\documentclass[pdflatex,sn-mathphys-num]{sn-jnl}

\usepackage[T1]{fontenc}
\usepackage[utf8]{inputenc}
\usepackage{lmodern}
\usepackage{graphicx}
\usepackage{xcolor}
\usepackage{amsmath,amssymb,amsfonts}
\usepackage{array,tabularx,booktabs,multirow,makecell}
\usepackage{enumitem}
\usepackage{rotating}
\usepackage{placeins}
\usepackage{xurl}
\usepackage{microtype}
\hypersetup{hidelinks}
\begin{document}

\title[6G Capabilities and Software-Business Value]{From 6G Capabilities to Software-Business Value: A Multivocal Review of Characteristics, Opportunities, Challenges}

\author*[1]{\fnm{Muhammad} \sur{Zohaib}}\email{muhammad.zohaib@lut.fi}
\author[1]{\fnm{Muhammad Azeem} \sur{Akbar}}
\author[1]{\fnm{Sami} \sur{Hyrynsalmi}}
\author[1]{\fnm{Muhammad} \sur{Hamza}}

\affil[1]{\orgname{Lappeenranta--Lahti University of Technology LUT},
\orgaddress{\city{Lappeenranta}, \postcode{53850}, \country{Finland}}}

\abstract{Sixth-generation (6G) technology is emerging as a transformative telecommunications paradigm capable of enabling intelligent, autonomous, and software-defined ecosystems. These developments could transform how software firms create, deliver, and capture value. Nevertheless, knowledge of the characteristics, business opportunities, and challenges of 6G software businesses remains fragmented across academic and practitioner literature. To address this gap, we conducted a Multivocal Literature Review of 133 primary sources (102 peer-reviewed studies and 31 grey literature sources) to examine the emerging 6G software-business landscape from the perspective of software-producing organizations. Through thematic mapping, we identified 18 characteristics across five higher-order themes, 32 business opportunities across eight themes, and 19 challenges and constraints across six themes. To assess classification reliability, three independent raters evaluated the factor-to-theme categorization using Fleiss' kappa, yielding substantial to almost perfect agreement across the three research questions ($\kappa = 0.744$--$0.832$). The findings highlight that value creation in 6G software businesses depends not only on technological capabilities but also on firms' ability to convert these capabilities into viable services and compelling business opportunities while addressing technological and ecosystem integration complexity, organizational, economic, and regulatory resistance, and security and sustainability constraints. Synthesizing across the research questions, we show that characteristics, opportunities, and challenges constitute interdependent dimensions of the emerging 6G software-business landscape. Integrating academic and practitioner evidence, the study establishes a rigorous, traceable foundation for understanding how software firms can create, deliver, and capture value within 6G ecosystems and a basis for future empirical research as 6G technologies, standards, and markets evolve.}


\maketitle

\section{Introduction}
Sixth-generation (6G) technology is expected to extend digital connectivity
beyond the capabilities of fifth-generation (5G) networks through increasingly
intelligent, software-defined, and distributed communication environments
\cite{ref1}.
While 5G primarily advances mobile broadband, low-latency communication,
and network reliability, 6G is envisioned to integrate artificial intelligence
(AI), machine learning (ML), edge computing, and advanced communication
technologies to support robust network autonomy and adaptability
\cite{ref2,ref3}. These developments
facilitate real-time and highly reliable communication across emerging digital
environments in which humans, machines, applications, and intelligent infrastructures
interact with increasing interdependence \cite{ref4}. Anticipated
capabilities include terabit-level data rates, intelligent spectrum management,
advanced edge intelligence, integrated sensing--communication--computation,
and satellite--ground coordination, which further illustrate the technological
transition toward highly programmable and software-intensive network environments
\cite{ref5,ref6,ref7,ref8,ref9}.

Moreover, the implications of this transition extend beyond network performance.
Emerging 6G capabilities support novel software-enabled
services in autonomous mobility, smart cities, immersive environments, healthcare,
industrial automation, and other data-intensive domains
\cite{ref2,ref10,ref11}. Realization of these
opportunities, however, requires more than access to advanced communication
technologies. Software-producing organizations must integrate connectivity,
computing, intelligence, data, and service orchestration across heterogeneous
infrastructures and multi-actor ecosystems \cite{ref12}.
Consequently, our assessment of the
transition toward 6G includes not only technological
feasibility but also the capacity of software firms to translate emerging
capabilities into viable services, business opportunities, and mechanisms for
value creation and capture.

In this study, we define 6G software businesses, the focal unit of analysis, as software-producing
organizations and platforms that design, develop, operate, or support software
systems, applications, services, and network functions within 6G environments.
Such organizations participate in ecosystems comprising the following actors: network
operators, infrastructure providers, cloud and edge providers, platform actors,
vertical-industry organizations, regulators, and end users
\cite{ref13,ref14,ref15}. Their potential roles extend from
software-defined networking and intelligent resource
orchestration to analytics, security, service platforms, and industry-specific
applications. Moreover, the commercial realization of these capabilities may
be influenced by interoperability, infrastructure readiness, security and privacy,
regulatory requirements, sustainability, operational complexity, ecosystem
dependencies, and uncertainty surrounding emerging markets and standards
\cite{ref16,ref17,ref18}.

Furthermore, existing secondary studies provide a substantial
knowledge base across multiple dimensions of 6G: Lu~\cite{ref17}
synthesizes technologies, scenarios, and challenges; Bhat~\cite{ref18}
analyzes the broader 6G ecosystem, technologies, business models, and sustainability;
Porambage~\cite{ref16} addresses security and privacy; and Al-Ansi~\cite{ref19}
examines intelligent edge computing together with use cases and
market drivers. More recent reviews extend the business-oriented
perspective: Singh~\cite{ref20} consider technology integration, implementation
challenges, and network monetization, while Pervaiz~\cite{ref21} examine
telecommunications entrepreneurship, business-model innovation, ecosystem
conditions, and emerging use cases. Collectively, these studies demonstrate
that the technological and business implications of 6G are increasingly being
considered together.

Nevertheless, we identified a gap in the existing literature. Existing reviews address
technology, network operator, security, edge computing, monetization,
and telecommunications entrepreneurship, but do not explicitly
examine software-producing organizations as the focal unit of analysis.
Moreover, the evidence remains dispersed across a substantial body of studies addressing
technological capabilities, commercialization opportunities, ecosystem
conditions, and challenge categories. Consequently, a consolidated understanding remains limited of how the characteristics associated with 6G
software businesses relate to the business opportunities they enable and to
the technical and non-technical conditions that constrain their development,
adoption, and scaling. This gap is particularly salient within an emerging,
pre-standardization domain, where practitioner and industry evidence does
complement the peer-reviewed literature in understanding evolving technologies,
markets, and ecosystem arrangements.

To address this gap, we conducted a Multivocal Literature Review (MLR)
in accordance with established Evidence-Based Software Engineering (EBSE)
and MLR guidelines
\cite{ref22,ref23,ref24}. By integrating peer-reviewed research with grey literature, the
review synthesized both academic and practitioner-oriented evidence concerning
the emerging 6G software-business landscape. Rather than proposing
new technologies or forecasting specific market outcomes, we systematically examined how existing evidence characterizes the capabilities,
opportunities, and constraints associated with software-producing organizations
in emerging 6G environments. Accordingly, we formulated the following
research questions:

\begin{enumerate}
    \item RQ1: What are the primary characteristics of 6G software businesses
    reported in the literature?
    \item RQ2: What business opportunities are associated with 6G software
    businesses?
    \item RQ3: What challenges and limitations influence the development and
    adoption of 6G software businesses?
\end{enumerate}

We organized the thematic synthesis of the evidence according to these three
complementary dimensions and subsequently examined the relationships among these dimensions
to provide an integrated understanding of software capabilities, potential
mechanisms for value creation and capture, and the conditions influencing
their realization. To assess the reliability of the resulting factor-to-theme
categorization, we commissioned an independent assessment by three additional
authors who applied Fleiss' kappa \cite{ref25}. Agreement was evaluated separately
for the three research questions and ranged ($\kappa = 0.744$--$0.832$), clearly indicating
substantial to almost perfect agreement according to the widely accepted
interpretation criteria \cite{ref26}.

We identified four primary contributions: First, we provided a multivocal
synthesis centered specifically on 6G software businesses, identifying 69 factors
comprising 18 characteristics, 32 business opportunities, and 19 challenges
and limitations, organized into 19 higher-order themes. Second, we bridged
these three dimensions to examine how emerging software capabilities relate to
potential mechanisms for value creation and capture, as well as the conditions
that constrain their realization. Third, we integrated peer-reviewed
and practitioner-oriented evidence to provide a comprehensive perspective on an emerging
technological and business landscape in which industry developments may
progress alongside formal research. Finally, the resulting structured and traceable evidence base provides a methodological foundation for
empirical investigation and for evaluating how software firms undertake
the development, commercialization, and scaling of 6G-enabled services as
technologies, standards, ecosystems, and markets mature.

The remainder of this paper is organized as follows.
Section~\ref{sec:background} (Background) introduces the
technological and business context of 6G software businesses.
Section~\ref{sec:methodology} (Methodology) describes the
Multivocal Literature Review methodology and review protocol.
Section~\ref{sec:results} (Results and Findings) presents
the demographic and thematic findings addressing the three
research questions. Section~\ref{sec:discussion} (Discussion) interprets the integrated
framework and its implications. Section~\ref{sec:validity} discusses
threats to validity, and Section~\ref{sec:implications} presents the
study implications and future work. Section~\ref{sec:conclusion}
concludes the paper and outlines directions for future research.
\section{Background}
\label{sec:background}
This section describes the background of software-enabled 6G systems and businesses. It covers the evolution toward software-centric wireless networks, key 6G technologies, the emerging software business ecosystem, major industry applications, and the main technical and non-technical challenges.
\subsection{Evolution of Wireless Networks Toward Software-Centric 6G Systems}
The advancement of wireless networks has been evolutionary, with successive generations introducing new capabilities in response to changing communication and service requirements. Early generations primarily supported voice communication, while later generations enabled mobile services, broadband communication, and integration with the Internet of Things (IoT) \cite{ref27}. 5G extended these capabilities through enhanced mobile broadband, low-latency communication, and support for applications such as autonomous vehicles, industrial control, and smart cities. However, emerging service requirements continue to place greater demands on network scalability, adaptability, intelligence, and energy efficiency. In this context, 6G systems are expected to become increasingly software-centric and intelligent, with greater automation and reduced dependence on fixed hardware-based network functions \cite{ref28,ref29}.

Compared with previous generations, 6G is envisioned to further shift network operation toward intelligent and software-controlled architectures \cite{ref30}. It is expected to integrate AI and ML across network and edge environments together with emerging communication and computing technologies such as Terahertz (THz) transmission and quantum communication \cite{ref13,ref31,ref32}. These developments may enable closer integration between physical and digital environments and support emerging applications in healthcare, autonomous transportation, intelligent manufacturing, and immersive digital communication \cite{ref32}.

This transition increases the role of software in network operation and service delivery, providing a technological foundation for emerging 6G software businesses.

\subsection{Key Technological Enablers of 6G}

Recent research identifies AI-driven automation as one of the major technological enablers of 6G systems \cite{ref2,ref33}. AI is expected to support functions such as resource allocation, self-healing, spectrum optimization, and security, contributing to more adaptive and flexible network architectures \cite{ref32}. This increasing integration of intelligence also creates a need for software systems that can respond dynamically to variations in network scale, conditions, and service requirements. Quantum communication represents another emerging direction for strengthening communication security in future networks. In particular, quantum key distribution (QKD) has been investigated as a mechanism for secure key exchange in the presence of emerging quantum-computing threats \cite{ref3,ref16}. Such capabilities may be relevant to security-sensitive domains, including military communications, financial services, and other applications requiring strong privacy protection \cite{ref34}. Blockchain-assisted and trust-aware approaches have also been investigated to support decentralized coordination, security, and trust in emerging 6G environments \cite{ref35}.

Edge computing and distributed intelligence are likewise expected to play an important role by moving computation and data processing closer to users and devices. This can reduce communication latency and improve responsiveness for applications such as autonomous driving, robotic-assisted surgery, and industrial automation \cite{ref19}. At the communication layer, THz spectrum is being investigated as a means of supporting extremely high data rates required by data-intensive applications such as holographic telepresence, extended reality (XR), and ultra-high-definition media services \cite{ref36,ref37}. However, THz communication also introduces challenges related to propagation loss, hardware requirements, and energy consumption, increasing the importance of effective network management and optimization \cite{ref38}.

These developments bring communication, computing, intelligence, and security capabilities closer together, requiring software to coordinate and adapt network functions and services.

\subsection{The 6G Software Business Ecosystem}

The increasing automation and integration of intelligence into 6G are expected to make software-producing organizations increasingly important within the emerging ecosystem. Beyond enabling connectivity, these actors may design, operate, and manage adaptive, AI-integrated, distributed, and software-defined network functions and services \cite{ref4,ref39}. From a software-business perspective, the 6G ecosystem involves multiple actors, including platform service providers, network service providers, vertical-domain service providers, infrastructure providers, regulators, and customers or end-users. These actors participate in the creation, delivery, commercialization, and scaling of software-enabled 6G services and capabilities \cite{ref14,ref40}. Such multi-actor environments require software mechanisms for coordinating complex networks, applications, and industry-specific services, particularly in mission-critical and industrial contexts \cite{ref41}. At the same time, software firms must operate within an environment shaped by security, privacy, compliance, and dependencies among multiple stakeholders \cite{ref42}.

The increasing role of AI-driven automation, edge computing, and decentralized trust mechanisms further illustrates the software-intensive nature of this ecosystem. AI-based automation may support self-healing, failure mitigation, anomaly detection, and continuous service management \cite{ref43}, while edge computing can support real-time processing for applications such as robotics, telemedicine, and smart energy systems \cite{ref30}. Blockchain-based approaches have also been investigated for decentralized identity, data integrity, and authentication in connected 6G environments \cite{ref16}. Their business implications, however, are addressed across separate technical, organizational, and policy-oriented studies.

\subsection{Industry Applications and Emerging Opportunities}

The development of intelligent and software-based 6G architectures is expected to support new applications across multiple industries through real-time data processing, automation, and advanced decision-making. In healthcare, high-quality and near-real-time data transmission may support remote diagnostics, robotic surgery, and patient monitoring \cite{ref44}. In transportation and smart-city environments, AI-enabled infrastructure and Vehicle-to-Everything (V2X) communication may support autonomous mobility, traffic coordination, road safety, and resource management \cite{ref45,ref46}. Similarly, 6G-enabled extended reality (XR) and AI-driven simulations may enable new forms of collaboration, training, and immersive digital services \cite{ref47}. Recent studies also examine integrated aerial, satellite, and ground networks, digital-twin environments, and edge-AI mechanisms for delivering responsive and adaptive services \cite{ref7,ref8}. These applications illustrate the range of potential software-enabled 6G services.

However, much of the existing research examines these application areas individually, often emphasizing technological feasibility, performance optimization, or specific system designs rather than their broader business implications. The links between enabling capabilities, potential business value, and implementation conditions are therefore not consistently examined together \cite{ref14,ref40}. These limitations become particularly relevant when moving from technically feasible applications toward scalable and sustainable software-based services, an issue considered further in the following subsection.

\subsection{Challenges and Fragmentation in 6G Software Business Research}

Despite the anticipated potential of 6G, existing studies identify security, trust, regulatory compliance, infrastructure costs, interoperability, and sustainability as important challenges to its development and adoption \cite{ref16,ref17}. The increasing reliance on AI-driven automation also raises concerns related to data integrity, adversarial attacks, and the governance of automated systems, creating additional security requirements for emerging network architectures \cite{ref16}. At the same time, evolving regulatory and standardization requirements concerning spectrum allocation, AI governance, and cross-jurisdictional data management may complicate large-scale deployment \cite{ref17}. Beyond these technical and regulatory concerns, research on innovation networks indicates that regional capabilities and collaboration structures can influence emerging 6G ecosystems, suggesting that technical feasibility alone is insufficient to explain adoption without considering organizational, economic, and policy conditions \cite{ref48,ref49}. Scalability and energy efficiency also remain relevant concerns for THz communication, AI-enabled edge computing, and other computationally intensive 6G environments \cite{ref17,ref50}.

Taken together, existing research provides substantial insight into individual 6G technologies, applications, and associated challenges. However, these perspectives remain dispersed across technical, business, organizational, and policy-oriented studies, with limited integration of practitioner-oriented evidence \cite{ref18}. This fragmentation makes it difficult to obtain a systematic understanding of how the characteristics of 6G software businesses relate to emerging business opportunities and the technical and non-technical conditions that may constrain their realization. A multivocal synthesis of academic and grey literature is therefore needed to consolidate these perspectives and provide a more comprehensive understanding of the emerging 6G software business landscape.
\section{Methodology}
\label{sec:methodology}

\begin{figure}[!htbp]
  \centering
  \includegraphics[width=\linewidth]{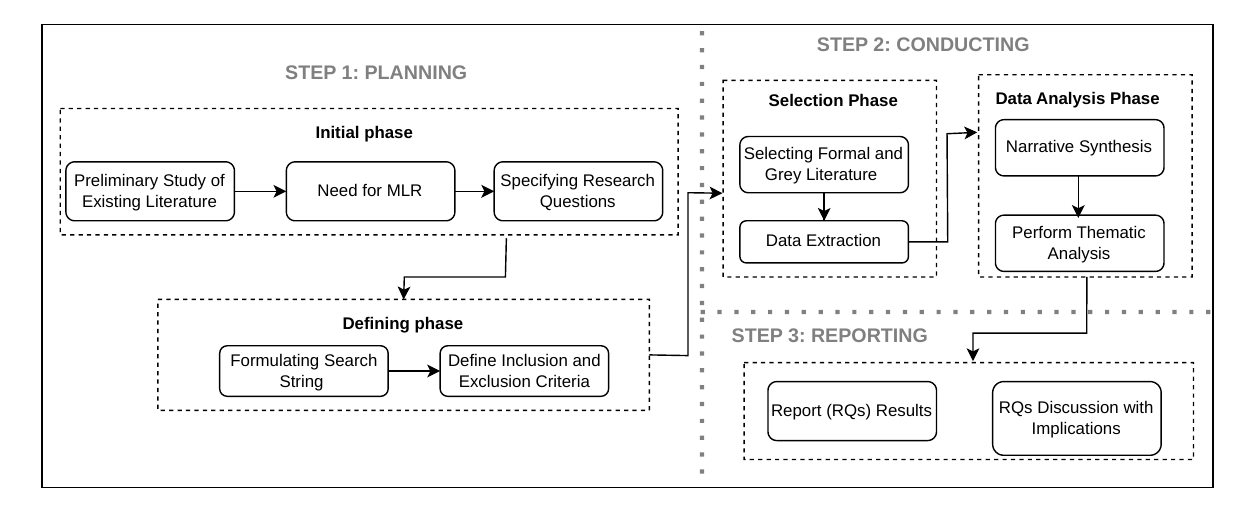}
  \caption{Research methodology}
  \label{fig:methodology}
\end{figure}

The study employs a Multivocal Literature Review (MLR) to systematically analyze the characteristics, opportunities, and challenges of 6G software businesses by synthesizing evidence from both formal scholarly literature and grey literature. The MLR methodology is particularly applicable to emerging and pre-standardization areas such as 6G, where industrial practice, ecosystem experimentation, and software-focused business innovation may progress faster than the peer-reviewed publication cycle \cite{ref22,ref51,ref52}. This combination captures both scholarly and practitioner perspectives. The methodological protocol follows Evidence-Based Software Engineering (EBSE) guidelines for secondary studies \cite{ref23,ref24} and complements them with the MLR guidelines proposed by Garousi et al. (2019). The review was structured into three stages: planning, conducting, and reporting, following established practices for systematic and multivocal reviews in software engineering \cite{ref22,ref24,ref53}. The overall methodological process is illustrated in Figure~\ref{fig:methodology}.

\subsection{Step 1: Planning}

The planning stage established the scope and design of the review by defining the rationale for the MLR, research questions, search strategy, and inclusion and exclusion criteria.
\subsubsection{MLR Preliminary Review and Rationale}

The exploratory review was conducted to develop familiarity with the domain and determine the maturity of research on 6G software businesses. According to the throw-away prototyping principle, the initial sources were used only to improve conceptual understanding and refine the search design and were therefore not automatically included in the final synthesis \cite{ref22,ref54}.

The structured checklist suggested by Garousi et al. (2019) was used to assess the rationale for including grey literature, as shown in Table~\ref{tab:gl_justification}. The assessment indicated that industry context and practitioner perspectives were important for understanding the emerging 6G software-business landscape, particularly because relevant evidence on ecosystem experimentation and emerging business models may not yet be adequately represented in formal scholarly literature. Accordingly, grey literature was incorporated alongside peer-reviewed literature through the MLR approach \cite{ref22,ref23,ref24}.

\renewcommand{\arraystretch}{1.2}
\begin{table}[!htbp]
\centering
\caption{Justification for Including Grey Literature \cite{ref22}}
\label{tab:gl_justification}
\footnotesize
\setlength{\tabcolsep}{5pt}
\begin{tabular}{@{}p{0.06\textwidth} p{0.69\textwidth} p{0.15\textwidth}@{}}
\hline
\textbf{\#} & \textbf{Question} & \textbf{Answer} \\ \hline
1 & Is the topic complex and not fully addressed by formal literature alone? & Yes \\ \hline
2 & Is there a shortage of formal literature on the topic? & No \\ \hline
3 & Is industry context essential for understanding the research topic? & Yes \\ \hline
4 & Can grey literature complement scientific findings with practitioner-oriented insights? & Yes \\ \hline
5 & Would combining academic and industry perspectives benefit the research? & Yes \\ \hline
6 & Is there significant industry interest in this topic? & Yes \\ \hline
\end{tabular}
\end{table}
\subsubsection{Research Questions}

The study selection, analysis, and synthesis were guided by a priori-defined research questions following EBSE guidelines \cite{ref23,ref24} and established MLR practices in software engineering \cite{ref22,ref55}. The review addresses three complementary dimensions of 6G software businesses: their characteristics, business opportunities, and challenges and limitations. The research questions, objectives, and rationale are presented in Table~\ref{tab:rqs}.

\renewcommand{\arraystretch}{1.3}
\begin{table}[!htbp]
\centering
\caption{Research Questions, Objectives, and Rationale}
\label{tab:rqs}
\footnotesize
\setlength{\tabcolsep}{5pt}
\begin{tabularx}{\textwidth}{
>{\raggedright\arraybackslash}p{0.24\textwidth}
>{\raggedright\arraybackslash}p{0.29\textwidth}
>{\raggedright\arraybackslash}X}
\hline
\textbf{Research Question (RQ)} & \textbf{Objective} & \textbf{Rationale} \\ \hline

RQ1: What are the primary characteristics of 6G software businesses reported in the literature? &
To identify the primary characteristics and software-oriented capabilities associated with 6G software businesses. &
6G is moving towards increasingly software-oriented architectures, creating new requirements and possibilities for software businesses and ecosystems \cite{ref39,ref52,ref56}. \\ \hline

RQ2: What business opportunities are associated with 6G software businesses? &
To identify the business opportunities, potential applications, and value-creation possibilities associated with 6G software businesses. &
Emerging 6G capabilities are expected to support new applications, services, business models, and ecosystem opportunities across different domains \cite{ref13,ref14,ref15,ref57}. \\ \hline

RQ3: What challenges and limitations influence the development and adoption of 6G software businesses? &
To identify the technical and non-technical challenges and limitations that may affect the development, adoption, and scaling of 6G software businesses. &
Existing literature identifies technical, security, economic, regulatory, and related challenges that may affect the development and adoption of 6G \cite{ref16,ref58,ref59,ref60}. \\

\hline
\end{tabularx}
\end{table}

These questions guided study selection, extraction, coding, and synthesis.
\subsubsection{Search Strategy}

To develop a systematic and focused search, the authors conducted a pilot search using IEEE Xplore and Google Scholar to refine the search terms relevant to 6G software businesses \cite{ref22,ref23,ref53}. The initial search showed that a broad search focused primarily on 6G retrieved substantially more records but with limited relevance to the business-oriented scope of this review, whereas only 62 papers identified through IEEE Xplore were specifically related to 6G software businesses. This observation supported the inclusion of business-oriented terms such as business model, ecosystem, platform, scenario, strategy, and value creation to improve the relevance of the search \cite{ref61,ref62,ref63}.

Based on the pilot search, the master search string was formulated by the first three authors and subsequently refined through group discussion \cite{ref22,ref53,ref64}. Search terms were combined using the Boolean operators ``AND'' and ``OR''. The master search string was then adapted to the syntax and search capabilities of seven academic databases while preserving its conceptual structure, following established systematic review guidelines \cite{ref22,ref23,ref53,ref55}. The final database-specific search strings and the number of retrieved records are presented in Table~\ref{tab:searchstrings} \cite{ref24}.

\begingroup
\footnotesize
\setlength{\tabcolsep}{5pt}
\renewcommand{\arraystretch}{1.15}

\begin{table}[!htbp]
\centering
\caption{Database-Specific Final Search Strings and Retrieved Records}
\label{tab:searchstrings}
\scriptsize
\setlength{\tabcolsep}{3pt}
\renewcommand{\arraystretch}{1.12}
\begin{tabular}{@{}>{\raggedright\arraybackslash}p{0.19\textwidth} >{\raggedright\arraybackslash}p{0.63\textwidth} >{\centering\arraybackslash}p{0.11\textwidth}@{}}
\hline
\textbf{Academic Database} & \textbf{Final Search String} & \textbf{Records} \\ \hline
IEEE Xplore &
("6G" OR "beyond 5G" OR B5G) AND
("business model" OR "business ecosystem" OR "business strategy"
OR "value creation" OR "business opportunity" OR "software business")
&
157\\ \hline
Scopus &
TITLE-ABS-KEY(("6G" OR "beyond 5G" OR B5G) AND
("business model" OR "business ecosystem" OR "business strategy"
OR "value creation" OR "business opportunity" OR "software business"))
&
152\\ \hline
Web of Science &
TS=(("6G" OR "beyond 5G" OR B5G) AND
("business model" OR "business ecosystem" OR "business strategy"
OR "value creation" OR "business opportunity" OR "software business"))
&
46\\ \hline
ACM Digital Library &
("6G" OR "beyond 5G" OR B5G) AND
("business model" OR "business ecosystem" OR "business strategy"
OR "value creation" OR "business opportunity" OR "software business")
&
410\\ \hline
SpringerLink &
"6G" AND
("business model" OR "business ecosystem" OR "business strategy"
OR "value creation" OR "software business")
&
433\\ \hline
ScienceDirect &
("6G" OR "beyond 5G") AND
("business model" OR "business ecosystem" OR "business strategy"
OR "value creation" OR "business opportunity")
&
651\\ \hline
Wiley Online Library &
("6G" OR "beyond 5G" OR B5G) AND
("business model" OR "business ecosystem" OR "business strategy"
OR "value creation" OR "business opportunity" OR "software business")
&
131\\ \hline
\textbf{Total} &
&
\textbf{1980}
\\ \hline
\end{tabular}
\end{table}
\endgroup
\subsubsection{Inclusion and Exclusion Criteria}

The inclusion and exclusion criteria were defined before the study-selection process, following the guidelines of Kitchenham et al. (2009) for formal literature and Garousi et al. (2019) for multivocal reviews. The criteria were developed by the first three authors to support consistent selection of studies relevant to the research questions.

For formal literature, studies were required to address the characteristics, opportunities, or challenges of 6G software businesses, be peer-reviewed and published in English, and have full text available. Duplicate studies, earlier versions of the same study, studies unrelated to the research questions, and studies focusing solely on hardware or physical-layer aspects were excluded. For grey literature, the criteria considered public availability, identifiable authorship, a minimum content length of 250 words, relevance to at least one research question, and relevant source types \cite{ref22,ref24}.

The criteria were applied sequentially during title, abstract, full-text, and quality screening, following established EBSE review practices \cite{ref24,ref53}. The inclusion and exclusion criteria for formal and grey literature are presented in Tables~\ref{tab:inc_exc} and~\ref{tab:greycriteria}, respectively, while the overall study-selection process is summarized in Figure~\ref{fig:prisma}.

\begingroup
\footnotesize
\setlength{\tabcolsep}{4pt}
\renewcommand{\arraystretch}{1.15}

\begin{table}[!htbp]
\centering
\caption{Inclusion and Exclusion Criteria for Formal Literature}
\label{tab:inc_exc}
\scriptsize
\setlength{\tabcolsep}{3pt}
\renewcommand{\arraystretch}{1.12}
\begin{tabular}{@{}>{\centering\arraybackslash}p{0.06\textwidth} >{\raggedright\arraybackslash}p{0.43\textwidth} >{\raggedright\arraybackslash}p{0.43\textwidth}@{}}
\hline
\textbf{No.} & \textbf{Inclusion Criteria} & \textbf{Exclusion Criteria} \\ \hline
1 &
Studies investigating the characteristics, opportunities, or challenges of 6G software businesses &

Secondary studies (e.g., systematic literature reviews, mapping studies, and review articles) were excluded from the primary evidence synthesis to avoid double counting.
\\ \hline

2 &
Peer-reviewed publications, including journal articles, conference papers, workshop papers, and symposium papers &
Earlier versions of the same study (only the most recent version retained)
\\ \hline

3 &
Studies for which the full text is available &
Studies not addressing any research question in the context of 6G software businesses
\\ \hline

4 &
Studies published in English &
Studies focusing solely on hardware or physical-layer aspects
\\
\end{tabular}
\end{table}
\endgroup

\begingroup
\footnotesize
\setlength{\tabcolsep}{4pt}
\renewcommand{\arraystretch}{1.15}

\begin{table}[!htbp]
\centering
\caption{Inclusion and Exclusion Criteria for Grey Literature}
\label{tab:greycriteria}
\scriptsize
\setlength{\tabcolsep}{3pt}
\renewcommand{\arraystretch}{1.12}
\begin{tabular}{@{}>{\centering\arraybackslash}p{0.06\textwidth} >{\raggedright\arraybackslash}p{0.43\textwidth} >{\raggedright\arraybackslash}p{0.43\textwidth}@{}}
\hline
\textbf{No.} & \textbf{Inclusion Criteria} & \textbf{Exclusion Criteria} \\ \hline
1 &
Publicly accessible sources &
Non-credible or unverifiable sources
\\ \hline

2 &
Clearly identifiable authorship (individual or organization) &
Anonymous or untraceable sources
\\ \hline

3 &
Content length sufficient for analysis (a minimum of 250 words) &
Sources with insufficient or superficial content
\\ \hline

4 &
Relevant to at least one research question &
Sources not contributing to any research question
\\ \hline

5 &
Industry reports, white papers, blogs, or technical documents relevant to 6G software businesses &
Non-English sources
\\
\end{tabular}
\end{table}

\endgroup
With the review scope, research questions, search strategy, and selection criteria established, the study proceeded to the conducting stage, where the literature search, study selection, quality assessment, data extraction, and thematic analysis were performed.
\subsection{Step 2: Conducting }

The conducting stage implemented the predefined review protocol through literature searching and study selection, quality assessment, data extraction and thematic analysis, followed by reliability and validation procedures.

\subsubsection{Literature Search and Study Selection}

\begin{figure}[!htbp]
  \centering
  \includegraphics[width=\linewidth]{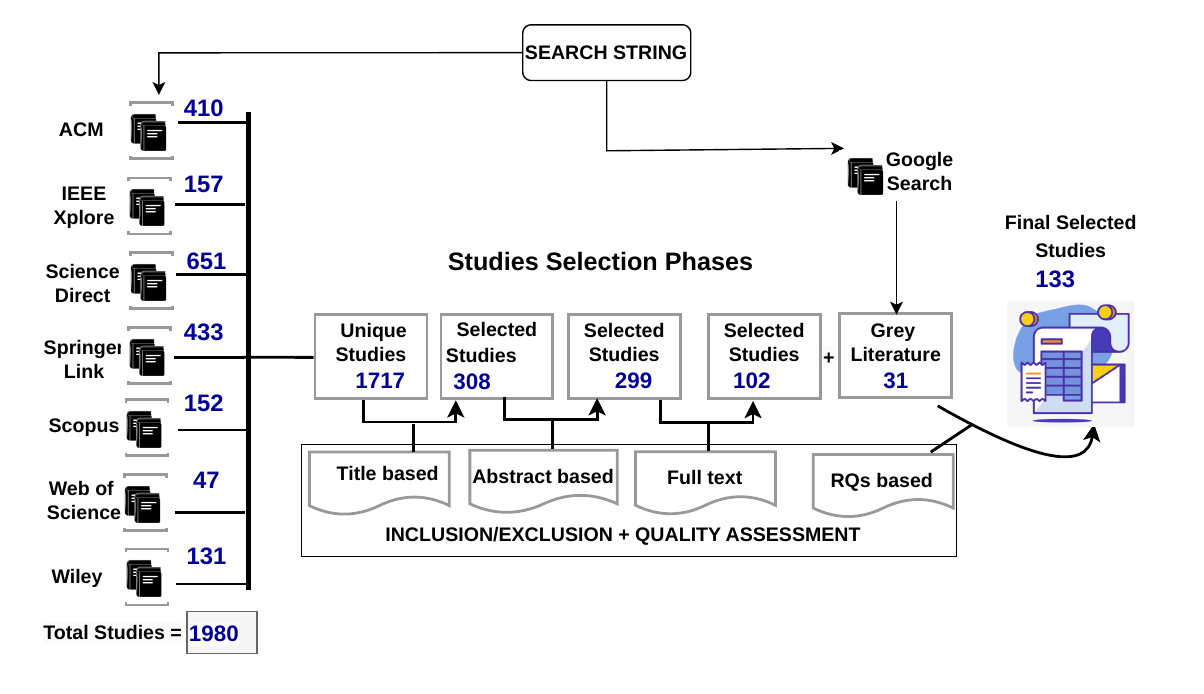}
  \caption{Study selection process}
  \label{fig:prisma}
\end{figure}

The literature search and study-selection process followed the predefined MLR and EBSE protocol \cite{ref22,ref23}. It comprised two complementary evidence streams: formal literature and grey literature. The predefined inclusion, exclusion, and quality-assessment criteria were applied through a multi-stage screening process. The overall study-selection process is illustrated in Figure~\ref{fig:prisma}.

For the formal literature, searches were conducted across the seven academic databases defined during the planning stage: IEEE Xplore, ACM Digital Library, Scopus, Web of Science, ScienceDirect, SpringerLink, and Wiley Online Library. The searches retrieved 1,980 records. After duplicate removal, 1,717 unique records remained for screening. Title-based screening reduced this set to 308 studies, followed by abstract-based screening, which retained 299 studies. Full-text assessment together with the predefined quality-assessment criteria resulted in 102 formal studies included in the final evidence base.

For the grey literature, Google Search was used to identify practitioner-oriented evidence in accordance with established MLR guidance \cite{ref22}. The search initially identified 150 potentially relevant grey-literature sources. These sources were assessed using the predefined grey-literature inclusion, exclusion, and quality criteria. Theoretical saturation and effort-bounded stopping criteria were applied during the search and screening process \cite{ref22}, resulting in 31 grey-literature sources being retained for the final synthesis.

Study selection was conducted by the first two authors using the predefined eligibility criteria. Screening decisions were compared at each stage, and disagreements were resolved through discussion. Where agreement could not be reached, the third author was consulted to make the final decision. The same procedure was applied when assessing the eligibility of grey-literature sources.

The final evidence base therefore comprised 133 primary sources, including 102 formal studies and 31 grey-literature sources. Complete bibliographic information, source identifiers, quality-assessment results, and study-selection traceability are provided in Supplementary Material S1~\cite{Zohaib2026Zenodo}.

\subsubsection{Literature Quality Assessment}

A Quality Evaluation Assessment (QEA) was conducted to assess the
relevance and methodological quality of the selected evidence following
established EBSE and MLR guidance
\cite{ref22,ref53,ref55}. For formal
literature, five quality criteria were applied to assess the
appropriateness of the research approach, relevance to the research
questions, clarity of the reported findings, relevance to the review
scope, and support for the reported findings
\cite{ref55,ref64,ref65}. Grey literature
was assessed using a 13-item checklist adapted from Garousi et al. (2019), covering source credibility, methodology,
objectivity, publication information, and contribution to the research
area.

All studies were assessed using a three-point scoring scheme
(Yes = 1, Partially = 0.5, No = 0). The total quality score for each
study was calculated relative to the maximum possible score of the
corresponding checklist. A predefined minimum quality score of 60\%
was used as an eligibility criterion for inclusion in the final
synthesis. Accordingly, formal studies were required to obtain at
least 3.0 out of 5 points, while grey-literature sources were required
to obtain at least 7.8 out of 13 points. The same scoring and decision
rules were applied consistently within each source type. The quality
assessment criteria and scoring schemes for formal and grey literature
are presented in Tables~\ref{tab:QEA_formal} and
\ref{tab:QEA_grey}, respectively, while the study-level quality
assessment scores are provided in Supplementary Material S1 at zenodo website\cite{Zohaib2026Zenodo} to support
transparency and traceability.


\renewcommand{\arraystretch}{1.2}
\begin{table}[htbp]
\centering
\caption{Quality evaluation assessment criteria for formal literature}
\label{tab:QEA_formal}
\footnotesize
\setlength{\tabcolsep}{4pt}
\begin{tabularx}{\textwidth}{
>{\raggedright\arraybackslash}p{0.13\textwidth}
>{\raggedright\arraybackslash}X
>{\raggedright\arraybackslash}p{0.25\textwidth}}
\hline
\textbf{QEA Questions} & \textbf{Checklist Questions} & \textbf{Scoring Scheme} \\
\hline

QEA1 &
Does the research approach used in the study adequately address the research questions? &
Yes = 1, Partial = 0.5, No = 0 \\ \hline

QEA2 &
Does the study discuss the characteristics, challenges, or opportunities of 6G software in a business context? &
Yes = 1, Partial = 0.5, No = 0 \\ \hline

QEA3 &
Are the reported findings clearly linked to the research questions? &
Yes = 1, Partial = 0.5, No = 0 \\ \hline

QEA4 &
Is the study relevant to the scope of this review (e.g., software business, future networks, or related domains)? &
Yes = 1, Partial = 0.5, No = 0 \\ \hline

QEA5 &
Are the findings clearly presented and well supported by evidence or analysis? &
Yes = 1, Partial = 0.5, No = 0 \\ \hline
\end{tabularx}
\end{table}


\renewcommand{\arraystretch}{1.2}
\begin{table}[htbp]
\centering
\caption{Quality evaluation assessment criteria for grey literature}
\label{tab:QEA_grey}
\footnotesize
\setlength{\tabcolsep}{4pt}
\begin{tabularx}{\textwidth}{
>{\centering\arraybackslash}p{0.18\textwidth}
>{\raggedright\arraybackslash}X
>{\raggedright\arraybackslash}p{0.25\textwidth}}
\hline
\textbf{Criteria Type} & \textbf{Checklist Questions (QEA)} & \textbf{Scoring Scheme} \\
\hline

Data Credibility &
QEA1: Does the published data originate from a reputable organization, or does the author have sufficient expertise in the field? &
Yes = 1, Partial = 0.5, No = 0 \\ \hline

\multirow{6}{0.18\textwidth}{\centering Methodology} &
QEA2: Is the purpose of the source clearly stated? &
Yes = 1, Partial = 0.5, No = 0 \\ \cmidrule{2-3}

& QEA3: Is a clear methodology described in the source? &
Yes = 1, Partial = 0.5, No = 0 \\ \cmidrule{2-3}

& QEA4: Are the references used in the source credible and up to date? &
Yes = 1, Partial = 0.5, No = 0 \\ \cmidrule{2-3}

& QEA5: Are any limitations explicitly stated? &
Yes = 1, Partial = 0.5, No = 0 \\ \cmidrule{2-3}

& QEA6: Does the work address a specific research question or objective? &
Yes = 1, Partial = 0.5, No = 0 \\ \cmidrule{2-3}

& QEA7: Does the work focus on a specific group, context, or case study? &
Yes = 1, Partial = 0.5, No = 0 \\ \hline

\multirow{3}{0.18\textwidth}{\centering Objectivity} &
QEA8: Is the presentation of the work balanced and unbiased? &
Yes = 1, Partial = 0.5, No = 0 \\ \cmidrule{2-3}

& QEA9: Are the statements presented objectively and supported by evidence rather than personal opinion? &
Yes = 1, Partial = 0.5, No = 0 \\ \cmidrule{2-3}

& QEA10: Are the conclusions supported by the presented data or evidence? &
Yes = 1, Partial = 0.5, No = 0 \\ \hline

Date &
QEA11: Is the publication date clearly specified? &
Yes = 1, Partial = 0.5, No = 0 \\ \hline

\multirow{2}{0.18\textwidth}{\centering Novelty} &
QEA12: Does the source provide distinctive or valuable contributions to the research area? &
Yes = 1, Partial = 0.5, No = 0 \\ \cmidrule{2-3}

& QEA13: Does the source strengthen or challenge existing knowledge or viewpoints? &
Yes = 1, Partial = 0.5, No = 0 \\

\hline
\end{tabularx}
\end{table}

\subsubsection{Data Extraction and Thematic Analysis}

Thematic analysis was used to extract and synthesize data from the 133
selected studies and identify patterns relevant to the research questions
\cite{ref66,ref67}. All studies were analyzed systematically,
and relevant data units were identified and coded. Related codes were
consolidated into factors, and conceptually related factors were subsequently
grouped into higher-order themes.

\begin{figure}[!htbp]
  \centering
  \includegraphics[width=\linewidth]{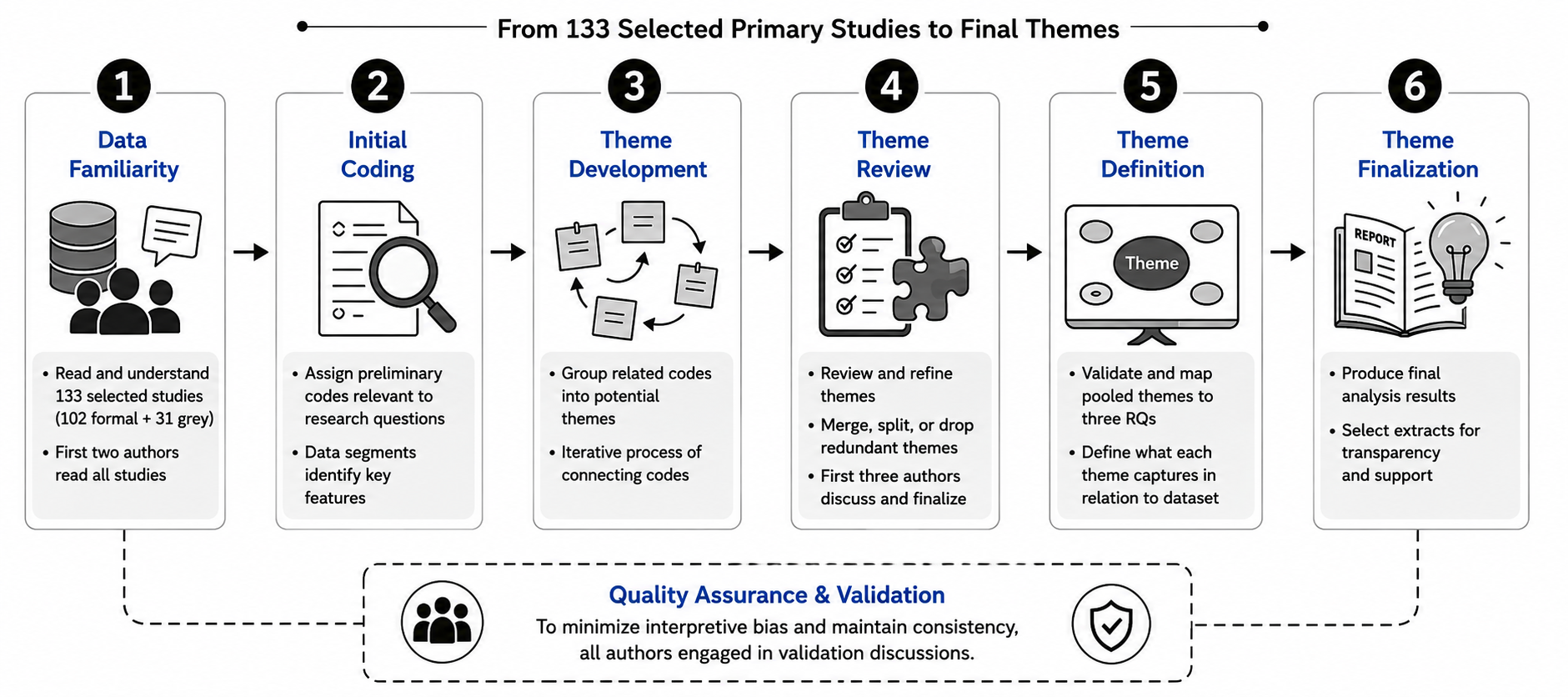}
  \caption{Thematic analysis and synthesis process}
  \label{fig:thematic}
\end{figure}

The analysis proceeded through six stages. First, during data
familiarization, the first two authors read the selected studies to develop
an understanding of the dataset. Second, during initial coding, relevant
data segments were assigned preliminary codes according to the research
questions. Third, during theme development, related codes and factors were
grouped into potential higher-order themes. Fourth, during theme review,
the first two authors reviewed and refined the emerging themes through
discussion, including the merging or removal of redundant themes. Fifth,
during theme definition, the themes were defined and mapped to the
corresponding research questions. Finally, during theme finalization, the
reviewed themes and their underlying factors were consolidated into the
initial thematic structure. Figure~\ref{fig:thematic} illustrates the
thematic-analysis process.

The reliability of the resulting factor-to-theme categorization was subsequently assessed by three independent raters who were not involved in developing the initial thematic classifications, as described in the following subsection. The study-to-factor coding and evidence mappings are provided in Supplementary Material S1 \cite{Zohaib2026Zenodo}.

\subsubsection{Reliability and Validation}

To assess the reliability of the thematic categorization, three independent raters classified the identified factors into the predefined higher-order themes. The raters were not involved in developing the initial factor-to-theme mappings. The assessment was conducted separately for each research question.
Specifically, the assessment included 18 characteristics across five
higher-order themes for RQ1, 32 opportunities across eight higher-order
themes for RQ2, and 19 challenges across six higher-order themes for RQ3.

During the independent assessment, each rater was provided with the
identified factors and the corresponding set of higher-order themes for
the relevant research question. The raters were blinded to the original
factor-to-theme assignments and to the classifications made by the other
raters. Inter-rater agreement was assessed using Fleiss' kappa ($\kappa$)
\cite{ref25}, which is suitable for evaluating agreement
among multiple raters assigning items to categorical groups. Agreement
was calculated separately for RQ1, RQ2, and RQ3 before any consensus
discussion.

The assessment resulted in Fleiss' kappa values of $\kappa = 0.813$ for
RQ1, $\kappa = 0.832$ for RQ2, and $\kappa = 0.744$ for RQ3. According
to commonly used interpretation criteria \cite{ref26},
these values indicate almost perfect agreement for RQ1 and RQ2 and
substantial agreement for RQ3. Full agreement was achieved for 14 of
the 18 factors in RQ1 (77.8\%), 25 of the 32 factors in RQ2 (78.1\%),
and 13 of the 19 factors in RQ3 (68.4\%). Overall, the three raters
reached full agreement on 52 of the 69 factors (75.4\%).

Following the independent assessment, the four manuscript authors reviewed the raters' classifications and resolved disagreements
through discussion to establish consensus on the final factor-to-theme assignments. The consensus process considered the original evidence, factor definitions, and independent classifications before finalizing the thematic structure.
Table~\ref{tab:inter_rater_reliability} summarizes the inter-rater agreement results, while the factor-level classifications and consensus assignments are provided in Supplementary Material S1~\cite{Zohaib2026Zenodo}.
.

To ensure consistent evidence traceability throughout the MLR, the 102
formal studies were assigned unique identifiers (FL-01--FL-102), while
the 31 grey-literature sources were assigned identifiers (GL-01--GL-31).
These identifiers are used to link the primary evidence to the extracted
characteristics, opportunities, and challenges and their corresponding
higher-order themes. The corresponding mappings are provided at \cite{Zohaib2026Zenodo} Supplementary Material S1.

\renewcommand{\arraystretch}{1.2}
\begin{table}[!htbp]
\centering
\caption{Inter-rater reliability of the factor-to-theme categorization}
\label{tab:inter_rater_reliability}
\footnotesize
\setlength{\tabcolsep}{6pt}
\renewcommand{\arraystretch}{1.18}
\begin{tabularx}{\linewidth}{@{}c c c c c >{\raggedright\arraybackslash}X@{}}
\hline
\textbf{RQ} &
\textbf{Factors} &
\textbf{Themes} &
\shortstack{\textbf{Full}\\\textbf{Agreement}} &
\shortstack{\textbf{Fleiss'}\\$\kappa$} &
\textbf{Interpretation} \\
\hline
RQ1 & 18 & 5 & 14 (77.8\%) & 0.813 & Almost perfect \\ \hline
RQ2 & 32 & 8 & 25 (78.1\%) & 0.832 & Almost perfect \\ \hline
RQ3 & 19 & 6 & 13 (68.4\%) & 0.744 & Substantial \\ \hline
Overall & 69 & 19 & 52 (75.4\%) & -- & RQ-specific $\kappa$ reported \\
\hline
\end{tabularx}
\end{table}
\subsection{Step 3: Reporting }

The final stage involved reporting the results of the review in accordance with the research questions. The selected studies were first examined through demographic analysis to characterize the evidence base. The thematic findings were then reported separately for RQ1, RQ2, and RQ3, covering the identified characteristics, business opportunities, and challenges and limitations of 6G software businesses. The relationships across these findings, together with their implications for research and practice, are further examined in Section~\ref{sec:discussion} (Discussion).

\section{Results and Findings}
\label{sec:results}
This section presents the findings of the MLR based on 133
primary studies, comprising 102 peer-reviewed publications
and 31 grey-literature sources. First, the Demographic
Analysis (Section~\ref{sec:demographics}) characterizes
the temporal, geographical, publication-channel, and
study-type distribution of the evidence base.

The thematic analysis then addresses RQ1: Characteristics
of 6G Software Businesses (Section~\ref{sec:rq1}),
RQ2: Business Opportunities (Section~\ref{sec:rq2}),
and RQ3: Challenges and Limitations
(Section~\ref{sec:rq3}). Overall, 69 factors were
identified 18 characteristics, 32 opportunities, and
19 challenges and synthesized into 19 higher-order
themes: five related to characteristics, eight to
opportunities, and six to challenges.

\subsection{Demographic Overview of the Reviewed Literature}
\label{sec:demographics}
The demographic analysis characterizes the evidence base underlying the thematic findings by examining publication trends, geographical distribution, publication channels, and study types. This provides context for interpreting the scope, maturity, and composition of the literature from which the RQ1--RQ3 findings were derived.

\subsubsection{Temporal Analysis of 6G Publication}
\begin{figure}[!htbp]
  \centering
  \includegraphics[width=\linewidth]{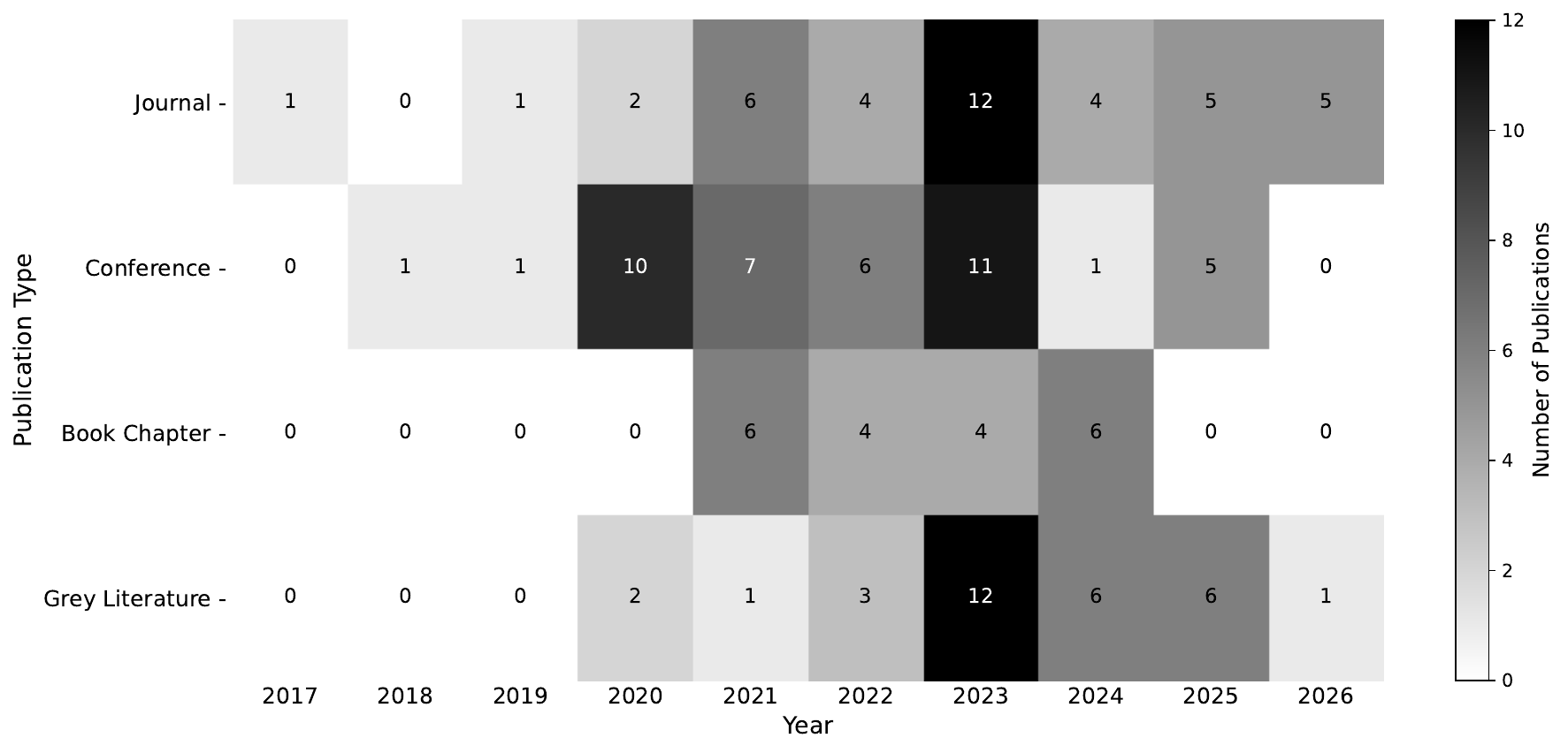}
  \caption{Year-wise Distribution of 6G Research Publications}
  \label{fig:yearwise}
\end{figure}

The 133 studies span the period from 2017 to 2026 across four publication channels (Figure~\ref{fig:yearwise}). Publication activity was limited during 2017--2019, with one study in 2017, one in 2018, and two in 2019. The volume increased substantially from 2020, with 14 publications, followed by 20 in 2021 and 17 in 2022. Output peaked in 2023 with 39 publications, representing 29.3\% of the entire corpus, with contributions appearing across all four publication channels. The number subsequently declined to 17 publications in 2024 and 16 in 2025. The 2026 count should be interpreted as partial because the search covered only part of that year.
Across publication channels, conference papers (n=42) and journal articles (n=40) were dominant, together accounting for 82 of the 133 studies (61.7\%). Grey literature contributed 31 sources (23.3\%) and reached its highest annual count in 2023, while book chapters accounted for 20 studies (15.0\%). Overall, the temporal distribution indicates that the evidence base expanded considerably after 2019, with its highest concentration occurring in 2023.

\subsubsection{Geographic Distribution of 6G Research Contributions}
\begin{figure}[!htbp]
  \centering
  \includegraphics[width=\linewidth]{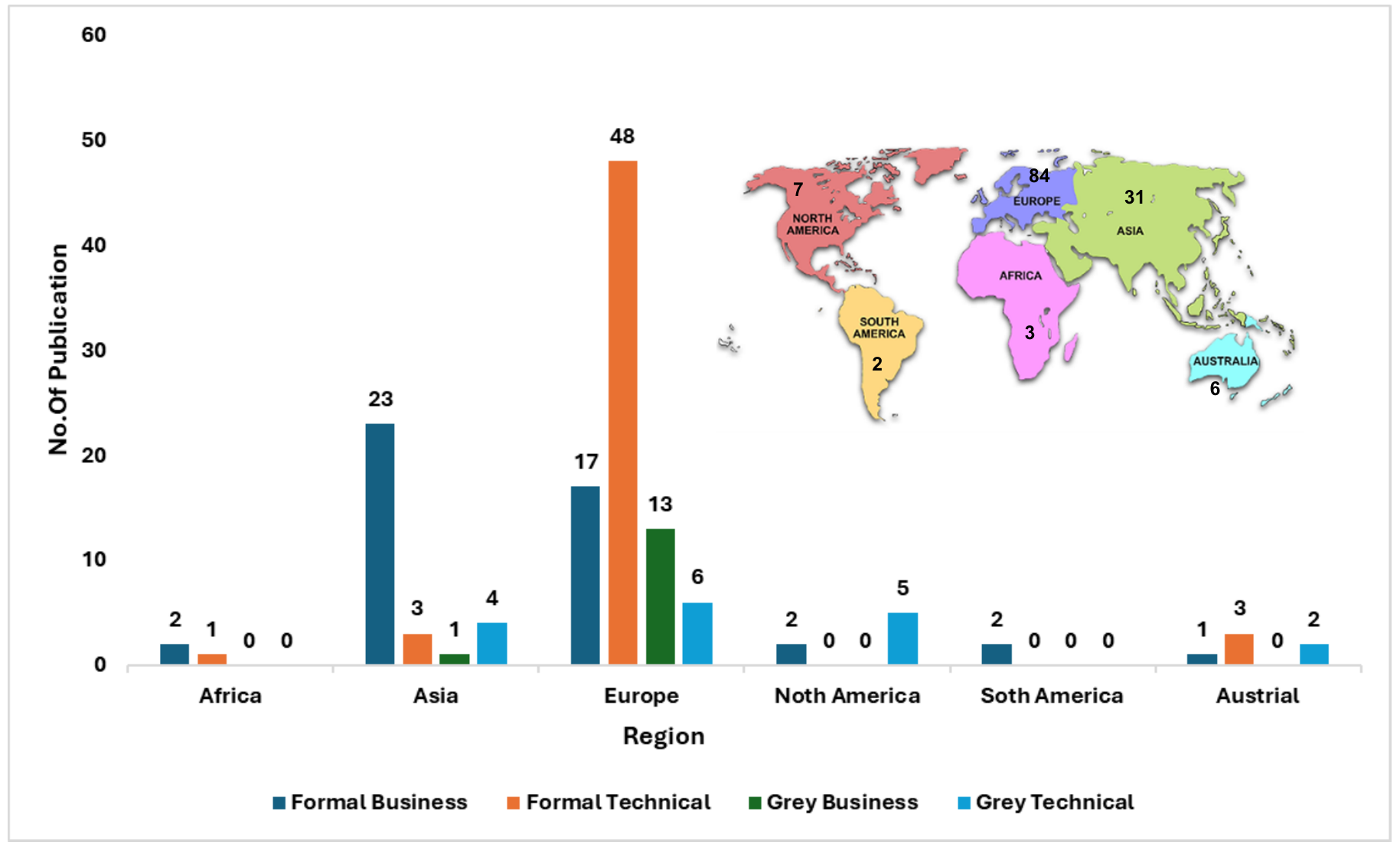}
  \caption{Geographic Distribution of 6G Research Publications}
  \label{fig:geo}
\end{figure}

Figure~\ref{fig:geo} presents the geographical distribution of the selected studies across formal and grey literature and business- and technical-oriented contributions. Europe represents the largest share of the evidence base, with particularly strong representation in formal technical literature. Asia also contributes substantially, whereas North America and Australia account for smaller shares of the corpus. Africa and South America are represented only marginally.

The distribution therefore reveals a geographically uneven evidence base, with the reviewed literature concentrated primarily in Europe and Asia. This imbalance should be considered when interpreting the thematic findings, as the identified characteristics, opportunities, and challenges may disproportionately reflect the technological, industrial, and institutional contexts represented in these regions.

\subsubsection{Distribution of Published Research Types in 6G Studies}
\begin{figure}[!htbp]
  \centering
  \includegraphics[width=\linewidth]{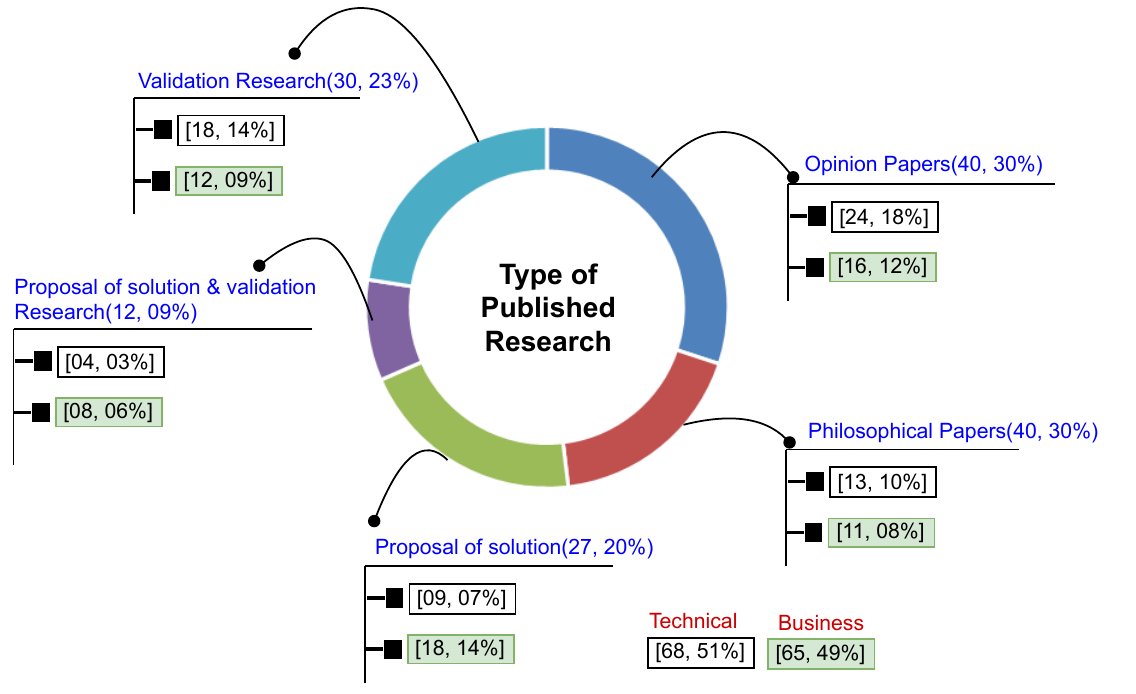}
  \caption{Distribution of 6G Research by Study Type}
  \label{fig:studytype}
\end{figure}

Figure~\ref{fig:studytype} shows the distribution of the selected literature by study type. Opinion and philosophical studies constitute a substantial part of the evidence base, reflecting the exploratory nature of research in the still-emerging 6G domain. Proposal-oriented studies further demonstrate considerable attention to prospective solutions and applications, whereas comparatively fewer studies provide empirical validation. This distribution indicates that the current evidence base remains predominantly conceptual and forward-looking rather than empirically mature.
\par\medskip
\noindent\rule{\columnwidth}{0.4pt}
\vspace{3pt}

\noindent\textbf{Key Takeaways (Demographic Overview)}

\begin{enumerate}
    \item The final evidence base comprises 133 primary sources, including 102 formal studies and 31 grey-literature sources, indicating substantial academic and practitioner interest in the emerging 6G software-business domain.

    \item Publication activity increased considerably after 2019 and reached its highest level in 2023, while the 2026 count represents only a partial publication year.

    \item The literature is geographically concentrated primarily in Europe and Asia, with comparatively limited contributions from Africa, South America, North America, and Australia.

    \item Conference papers and journal articles constitute the dominant publication channels, while grey literature provides an additional practitioner-oriented perspective on the emerging 6G software-business landscape.

    \item The evidence base remains largely conceptual and forward-looking, with opinion, philosophical, and proposal-oriented studies more prominent than empirically validated research.
\end{enumerate}

\vspace{1pt}
\noindent\rule{\columnwidth}{0.4pt}
\par\medskip

\subsection{Thematic Analysis: Characteristics of 6G Software Businesses}
\label{sec:rq1}

\begin{sidewaysfigure}[p]
  \centering
  \includegraphics[width=\textheight,height=0.84\textwidth,keepaspectratio]{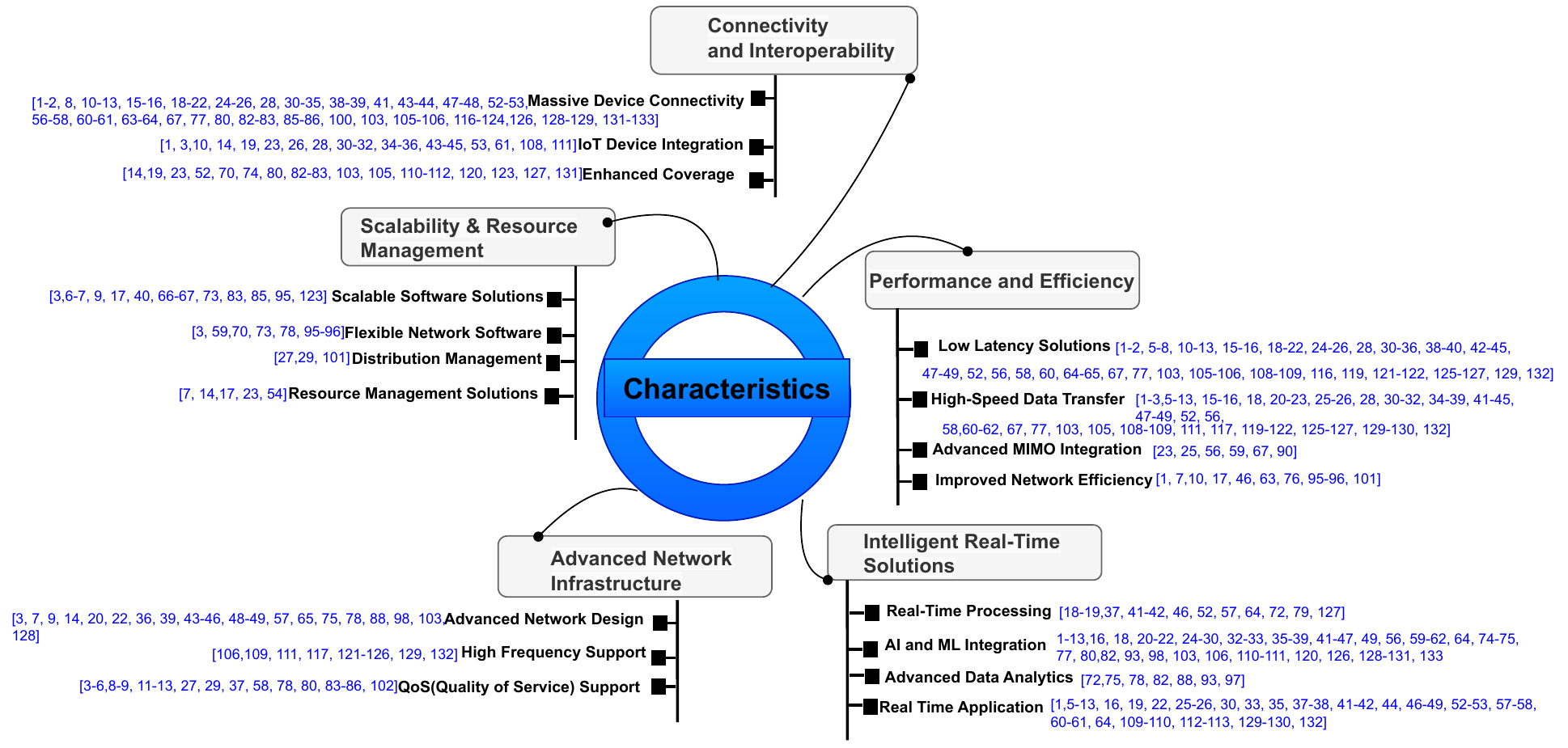}
  \caption{Primary characteristics of 6G software}
  \label{fig:char}
\end{sidewaysfigure}

The 18 characteristics identified for RQ1 were organized into five higher-order themes: Performance and Efficiency, Real-Time Intelligence, Scalability and Resource Management, Connectivity and Interoperability, and Advanced Network Infrastructure. Figure~\ref{fig:char} summarizes their distribution across these themes.

\subsubsection{Performance and Efficiency}

The fundamental component providing 6G with (Chac1) low-latency solutions (FL-01, FL-10, FL-20) pertains to time-critical software service applications, for example, autonomous driving, industrial automation, and real time financial systems (FL-14, FL-18, FL-20--FL-23). But latency, in isolation, is not enough to guarantee the business case for (Chac2) high-speed data transfer (FL-01, FL-08--FL-11, FL-16). This feature may allow software businesses to provide bandwidth-intensive services like immersive AR/VR applications, real time cloud services, and data streaming at terabit-per-second (Tbps) rates (FL-01, FL-08--FL-11, FL-42, FL-61; GL-27--GL-28). While these benefits expand the scope of services, they also increase operational complexity and the risk of network congestion. Therefore, (Chac3) network efficiency will need to be improved. This will require AI network optimization, self-healing, and dynamic resource allocation to maintain service quality and operational stability (FL-46, FL-63, FL-76, FL-95--FL-96, FL-101). Furthermore, (Chac4) new MIMO technologies will further assist with the effective and efficient use of the available frequency spectrum and data throughput, via massive MIMO, intelligent beamforming, and reconfigurable intelligent surfaces (RIS) for effective, efficient, scalable, and reliable service provision (FL-23, FL-25, FL-90).

\subsubsection{Solutions in real-time intelligence}

Due to the reliance on software, 6G demands the design of systems that can react to changes in the operational environment. (Chac5) Processing data in real time means that analytics and decisions can occur with minimal delay which is very important in areas like industrial control, finance, and remote healthcare (FL-18--FL-19, FL-37, FL-41--FL-42, FL-72, FL-79; GL-25). In 6G, the volume and the speed of data make (Chac6) the integration of AI and ML increasingly important as manual control becomes less feasible. AI-led intelligence facilitates self-optimizing networks, customizable security systems and the continuous adjustment of operational performance across the networks (FL-16, FL-18, FL-24--FL-30, FL-41--FL-47; GL-26--GL-29, GL-31). All these capabilities are enhanced by (Chac7) advanced data analytics, which supports predictive maintenance, anomaly detection, and proactive service management, potentially reducing the operational risks and downtimes (FL-72, FL-75, FL-78, FL-82, FL-88, FL-93, FL-97). All these capabilities enhance (Chac8) real-time applications, where software systems are able to autonomous decision-making in smart cities, collaborative environments, and industrial automation (FL-01, FL-22, FL-30, FL-33, FL-41--FL-42, FL-47, FL-52--FL-53; GL-27--GL-28, GL-30).

\subsubsection{Scalability and Resource Management}

To facilitate large-scale ecosystem deployment, (Chac9) scalable software solutions help 6G systems be flexible and fit different demands by means of virtualization, cloud-native architecture, and software-defined networking (FL-06--FL-07, FL-40, FL-66--FL-67, FL-83, FL-95; GL-21). These features are enhanced by (Chac10) flexible network software, which fosters the cross industrial, organizational, and regulatory software interoperability (FL-03, FL-59, FL-70, FL-73, FL-78, FL-95--FL-96). Efficient operation relies on (Chac11) intelligent resource management, which dynamically distributes bandwidth, computational, and storage resources according to the real-time prevailing service demands (FL-07, FL-14, FL-17, FL-23, FL-54). In (Chac12) geographically distributed deployments, distribution management reduces congestion and improves resilience to high traffic by balancing workload distribution across different regions (FL-27, FL-29, FL-101).

\subsubsection{Connectivity and Interoperability}

6G software businesses span highly interconnected ecosystems. (Chac13) massive device connectivity provides the basis for the integration of billions of IoT and edge devices and the basis for the coordination of ecosystem and platform-based services (FL-01, FL-08, FL-10, FL-14; GL-01, GL-03--GL-04, GL-14--GL-22). In addition to volume of connectivity, (Chac14) IoT device integration enables cross-domain service innovation through secure and seamless interactions at all levels of heterogeneous platforms and technologies (FL-01, FL-10, FL-23, FL-26, FL-28, FL-30--FL-32, FL-34--FL-36, FL-61; GL-06, GL-09). These features are additionally enhanced by (Chac15) enhanced coverage which provides access in less densely populated/up to previously underserved regions and supports the wider implementation of 6G-enabled services (FL-14, FL-19, FL-80, FL-82--FL-83; GL-01, GL-08--GL-10, GL-29).

\subsubsection{Advanced Network Infrastructure}

The future of 6G software businesses hinges on (Chac16) network architecture focused on resilience, energy efficiency, and AI-based decentralized optimization (FL-03, FL-07, FL-09, FL-14, FL-20, FL-36, FL-39, FL-43--FL-46, FL-48--FL-49, FL-57, FL-65, FL-75, FL-78, FL-88, FL-98; GL-01, GL-26). (Chac17) High-frequency support, including THz bands, is both a promoter of ultra-high bandwidth services and a cause of architectural and operational challenges for software systems (GL-04, GL-19--GL-24, GL-27, GL-30). Lastly, (Chac18) Quality of Service (QoS) support, entails the use of network slicing and predictive resource provisioning, for these are the type of metrics that ensure reliability of service provision within the intricate 6G ecosystem, and therefore are the type of metrics that ensure mission-critical applications adhere to performance guarantees (FL-04--FL-06, FL-11--FL-12, FL-27, FL-29, FL-37, FL-58, FL-80, FL-83--FL-85, FL-102).

Overall, the identified characteristics represent recurring capabilities that shape how software-producing organizations design, operate, and scale software-enabled services in emerging 6G ecosystems.
\par\medskip
\noindent\rule{\columnwidth}{0.4pt}
\vspace{3pt}

\noindent\textbf{Key Takeaways (RQ1)}

\begin{enumerate}
    \item Performance-related characteristics combine low latency, high-speed data transfer, network efficiency, and advanced MIMO capabilities to support demanding 6G software services.

    \item Real-time processing, AI/ML integration, advanced analytics, and real-time applications characterize the increasing role of intelligence and automation in 6G software environments.

    \item Scalable and flexible software, together with intelligent resource and distribution management, supports operation across changing service demands and distributed deployments.

    \item Massive connectivity, IoT integration, enhanced coverage, advanced network architecture, high-frequency support, and QoS characterize the heterogeneous infrastructure within which 6G software services operate.
\end{enumerate}

\vspace{1pt}
\noindent\rule{\columnwidth}{0.4pt}
\par\medskip

\subsection{Thematic Analysis: The Opportunities of 6G Software Businesses}
\label{sec:rq2}

\begin{sidewaysfigure}[p]
  \centering
  \includegraphics[width=\textheight,height=0.84\textwidth,keepaspectratio]{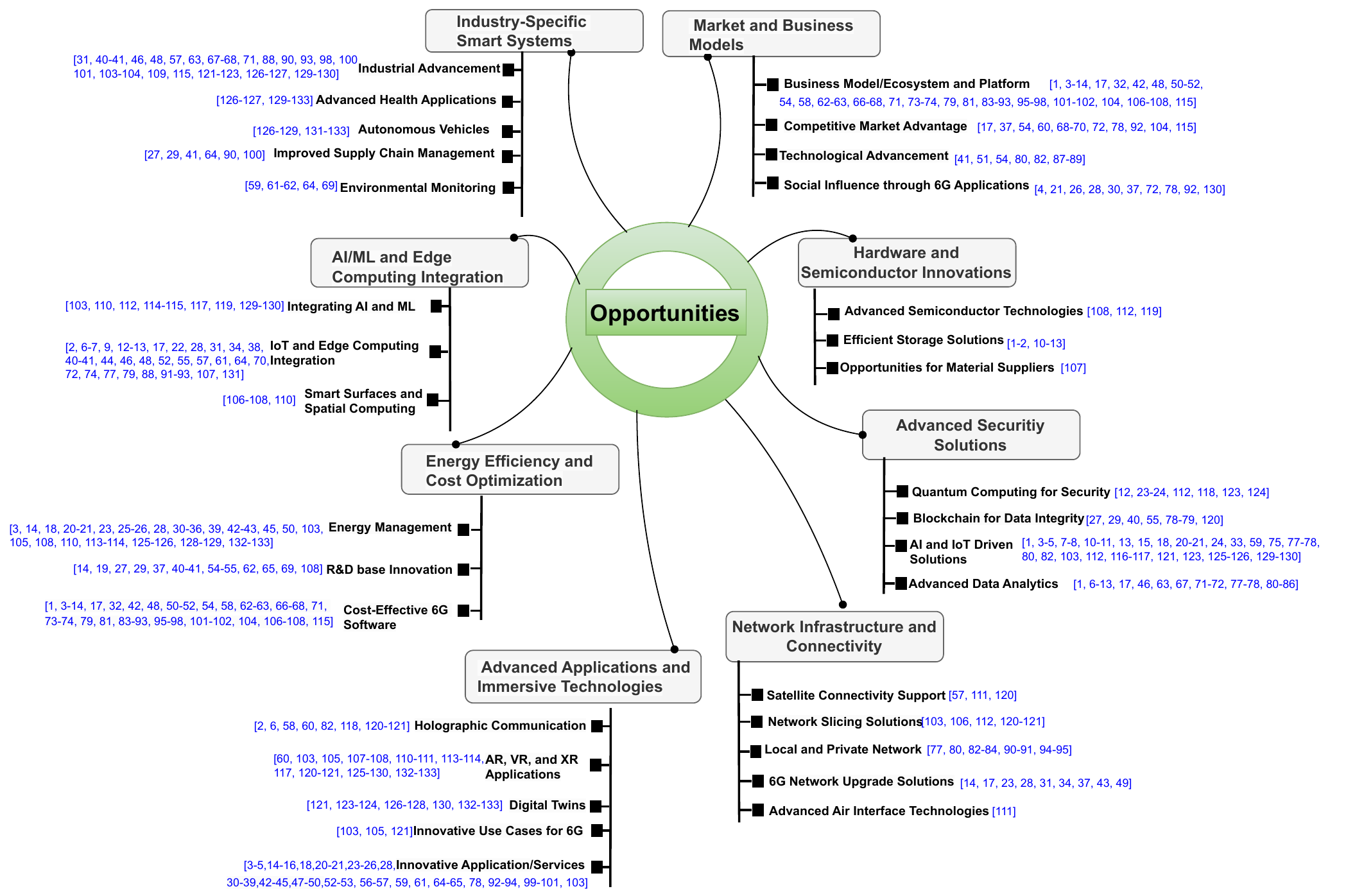}
  \caption{High Level 6G Software Business Opportunities}
  \label{fig:opps}
\end{sidewaysfigure}

The 32 opportunities identified for RQ2 were organized into eight higher-order themes: Security, Data Integrity, and Privacy; Network Infrastructure and Connectivity; Advanced Applications and Immersive Services; Sectoral Innovations; AI, ML, and Edge Computing Integration; Hardware and Semiconductor Innovations; Energy and Cost Optimization; and Business Models and Ecosystem Development. Figure~\ref{fig:opps} summarizes these opportunities across the eight themes.

\subsubsection{Security, Data Integrity, and Privacy: Trust-Centered Value Opportunities}

Trust is a fundamental opportunity for 6G software businesses in the context of highly distributed and highly autonomous environments. (Opp1) Quantum computing for cybersecurity is poised to foster sophisticated, more secure communications and improved resilience to cyberattacks, thus providing confidence in mission-critical digital services (FL-12, FL-23; GL-10, GL-16, GL-21--GL-22). (Opp2) Partnership with blockchain for secure data, integrity and service offers trust and transparency for non-participating entities in relation to the secure, tamper-proof, and transparent data management that supports inter-organizational safe transactional and cooperative processes (FL-27, FL-29, FL-40, FL-56, FL-78--FL-79; GL-18). Meanwhile, combined (Opp3) AI and IoT security solutions offer ongoing risk assessment, flexible defenses and immediate risk mitigation, operational risk reductions for software-based platforms (FL-01, FL-10, FL-12, FL-20--FL-21, FL-47, FL-82; GL-01, GL-10, GL-14--GL-15). From (Opp4) advanced data analytics further predictive security and preemptive control, offer scalable trust management across complex ecosystems (FL-01, FL-17, FL-46--FL-47, FL-63, FL-80--FL-86).

\subsubsection{Network Infrastructure and Connectivity: Scalable Service Enablement}

In business perspective, infrastructure related opportunities are focused on the expansion of services reach and flexible deployment model. (Opp5) The inclusion of satellite connectivity expands coverage of under-served areas, thus allowing software business to participate in inclusive digital markets and to offer worldwide services (FL-54--FL-55, FL-101; GL-01, GL-09, GL-18). At the service layer, (Opp6) network slicing solutions permit differentiated service offerings by customizing virtual network instances to fulfill application-specific requirements to enable high-value services and industry-specific value propositions (FL-66, FL-89; GL-01, GL-04, GL-10, GL-18--GL-19). In the same manner, (Opp7) local and private networks offer the opportunity to the vertical industries to implement the secure and tailored connectivity and thereby minimize the dependence on centralized infrastructures and promote the implementation of new B2B business models (FL-77, FL-80, FL-82--FL-84, FL-90--FL-91, FL-94--FL-95). In order to facilitate transition pathways, (Opp8) 6G network upgrade solutions will allow a progressive shift between the current systems, thus reducing the adoption barriers and protecting the investments done so far (FL-14, FL-17, FL-23, FL-28, FL-31, FL-34, FL-37, FL-43, FL-49, FL-96). The opportunities are supported with (Opp9) advanced air-interface technologies which enhance spectral efficiency and service reliability in dynamic operating conditions (FL-10, FL-58, FL-90; GL-09).

\subsubsection{Advanced Applications and Immersive Services: Differentiated Digital Offerings}

The 6G technology allows designing new application classes to enable software developers to develop a differentiated offering which goes beyond connectivity. (Opp10) Holographic communication enables immersive teamwork and telepresence, which may enable new value-added services in education, healthcare, and business teamwork (FL-06, FL-24, FL-43; GL-16, GL-18--GL-19). In the same way, (Opp11) applications based on augmented (AR), virtual (VR), and extended reality (XR) provide new services for immersive training, simulation, entertainment, and industrial operations, and utilize ultra-low latency and high-bandwidth networks to offer new value propositions (FL-33, FL-42--FL-43, FL-60; GL-01, GL-05--GL-06, GL-23--GL-28, GL-30--GL-31). (Opp12) The definition of digital twins involves ongoing evaluation and maximization of the performance of physical assets. This definition aids continuous, data driven decision making in smart cities, manufacturing, and management of infrastructures (FL-37, FL-43, FL-47; GL-19, GL-21--GL-22, GL-30--GL-31). Wider (Opp13) pioneering use cases and (Opp14) combined merged application services further may provide advantages by integrating AI, IoT, and blockchain for cross-domain interdisciplinary added value bundled use improvement of operational productivity and customization of services (Opp13: FL-20, FL-61--FL-62; GL-01, GL-03, GL-19; Opp14: FL-12, FL-14, FL-20--FL-21, FL-23--FL-26, FL-92--FL-94, FL-99--FL-101; GL-01).

\subsubsection{Innovations by sectors: Vertical Value Realization}

Numerous opportunities are interlaced within sectors where the 6G functionality and the specific domain needs are colliding/interlocking. (Opp15) industrial advancement enables automation, predictive maintenance, and real time optimization in manufacturing and logistics, fostering productivity enhancements and cost containment (FL-31, FL-40--FL-41, FL-47, FL-67--FL-68, FL-71, FL-90; GL-01, GL-02, GL-07, GL-13). In medicine, (Opp16) sophisticated health applications support remote diagnostics, assisted tele-surgery, and decision making (via AI) with the potential to improve service accessibility and quality (FL-42, FL-47; GL-24--GL-25, GL-27--GL-31). In the same vein, (Opp17) autonomous driving cars leverage on the real time data \& V2X (Vehicle-to-Everything) communications to deliver safer and more efficient mobility services (FL-39, FL-42, FL-47, FL-53; GL-24, GL-27). Besides, other emerging opportunities are found in (Opp18) refined supply chain analytics and blockchain integration to achieve transparency (FL-27, FL-29, FL-41, FL-64, FL-90, FL-100) and (Opp19) environmental compliance and monitoring to achieve sustainable services (FL-55, FL-69, FL-88, FL-95; GL-01).

\subsubsection{Integrating AI, ML, And Edge Computing: Operational Intelligence As A Service}

The core components of operational differentiation are edge computing and artificial intelligence/machine learning. (Opp20) AI-ML integration results in predictive analytics, autonomous control, and continuous optimization, thus allowing software companies to increase the delivery of intelligent services (FL-25, FL-47, FL-62; GL-01, GL-12--GL-13, GL-15--GL-17). At the same time, (Opp21) integration of the Internet of Things with edge computing enables localized processing, which can reduce latency and bandwidth dependence, which is essential to time-sensitive and mission-critical applications (FL-22, FL-28, FL-31, FL-34, FL-38, FL-40--FL-41, FL-61; GL-05, GL-29). The potential of contextual and intelligent interaction is further facilitated by emerging [Opp22] intelligent surfaces and spatial computing in the range of industrial and consumer applications (GL-04--GL-06, GL-08).

\subsubsection{Innovations in Hardware and Semiconductors: Software-Centric Innovations}

Most innovations are technological and hardware advances that create residual business opportunities. The implementation of AI processes and high-performance computing can be performed using (Opp23) advanced semiconductors in a highly scalable fashion, thus enabling the creation of software services (GL-06, GL-10, GL-17). Enhanced storage (Opp24) offers the prospect of eliminating the processing bottlenecks and speeding up solutions to data-intensive applications (FL-01--FL-02, FL-10--FL-13), and (Opp25) suppliers of materials to high-performance and low-energy components will be available to seal gaps (FL-03; GL-05).

\subsubsection{Energy and Cost Optimization: Sustainable Business Operations}

Issues related to energy consumption and costs are major potential opportunities for sustainable solutions. (Opp26) Energy-management systems, which use AI optimization along with the renewable resources, may reduce operating costs and environmental impacts (FL-03, FL-20--FL-21, FL-23, FL-28, FL-42--FL-43, FL-50; GL-01, GL-03, GL-26--GL-27). Research-based cost reduction initiatives (Opp27) focus on modular architectures and smart automation in order to lower costs of deployment and maintenance (FL-14, FL-19, FL-27, FL-29, FL-40--FL-41, FL-65, FL-69; GL-06). This program is supported by (Opp28) cost-effective 6G software solutions which improve efficiency and scalability especially in cloud-native and open-source architectures (FL-14--FL-16, FL-28--FL-29, FL-34--FL-40, FL-47, FL-49, FL-51--FL-54, FL-87; GL-19, GL-22).

\subsubsection{Business Models and Ecosystem Development: Monetizing 6G Capabilities}

Value realization requires business model innovation and ecosystem orchestration. (Opp29) Model, ecosystem, and platform innovation seek to commercialize market, service platform, and subscription model creation of 6G services (FL-17, FL-54, FL-67, FL-71, FL-73--FL-74, FL-101--FL-102; GL-02, GL-13). (Opp30) Differentiation competitive market advantage is therefore captured in terms of AI-enabled business intelligence and decision-making based on data (FL-17, FL-52, FL-54, FL-68--FL-70, FL-73, FL-92; GL-02, GL-13). Opportunities to add value through (Opp31) technology, i.e. software-defined networking and automation, and (Opp32) social impact (influence) are created by the evolution of 6G applications in education, health, smart cities, and social urbanization (Opp31: FL-40--FL-41, FL-51, FL-54, FL-80, FL-82, FL-87--FL-89; Opp32: FL-04, FL-26, FL-28, FL-30, FL-72, FL-78, FL-92; GL-28).

Overall, these opportunities indicate that 6G software-business value creation depends on combining enabling capabilities into software-enabled services and business models. Their relationships with the characteristics and challenges identified in RQ1 and RQ3 are examined further in Section~\ref{sec:cross-rq-synthesis} (Cross-RQ Synthesis).
\par\medskip
\noindent\rule{\columnwidth}{0.4pt}
\vspace{3pt}

\noindent\textbf{Key Takeaways (RQ2)}

\begin{enumerate}
    \item Security, data integrity, privacy, and advanced analytics create trust-centered opportunities for software-enabled 6G services.

    \item Satellite connectivity, network slicing, private networks, network upgrades, and advanced air-interface technologies expand the infrastructure and service configurations available to software businesses.

    \item Immersive applications and sector-specific innovations extend 6G software opportunities across healthcare, mobility, industry, supply chains, environmental monitoring, and other application domains.

    \item AI/ML, edge computing, IoT integration, and intelligent surfaces support more localized, adaptive, and intelligent software services.

    \item Hardware, semiconductor, storage, material, energy, and cost-related opportunities provide supporting conditions for the development and scaling of software-enabled 6G services.

    \item Business-model, ecosystem, platform, technological, and social opportunities shape how 6G capabilities may be translated into market and service value.
\end{enumerate}

\vspace{1pt}
\noindent\rule{\columnwidth}{0.4pt}
\par\medskip

\subsection{Thematic Analysis: Challenges in 6G Software Businesses}
\label{sec:rq3}

\begin{sidewaysfigure}[p]
  \centering
  \includegraphics[width=\textheight,height=0.84\textwidth,keepaspectratio]{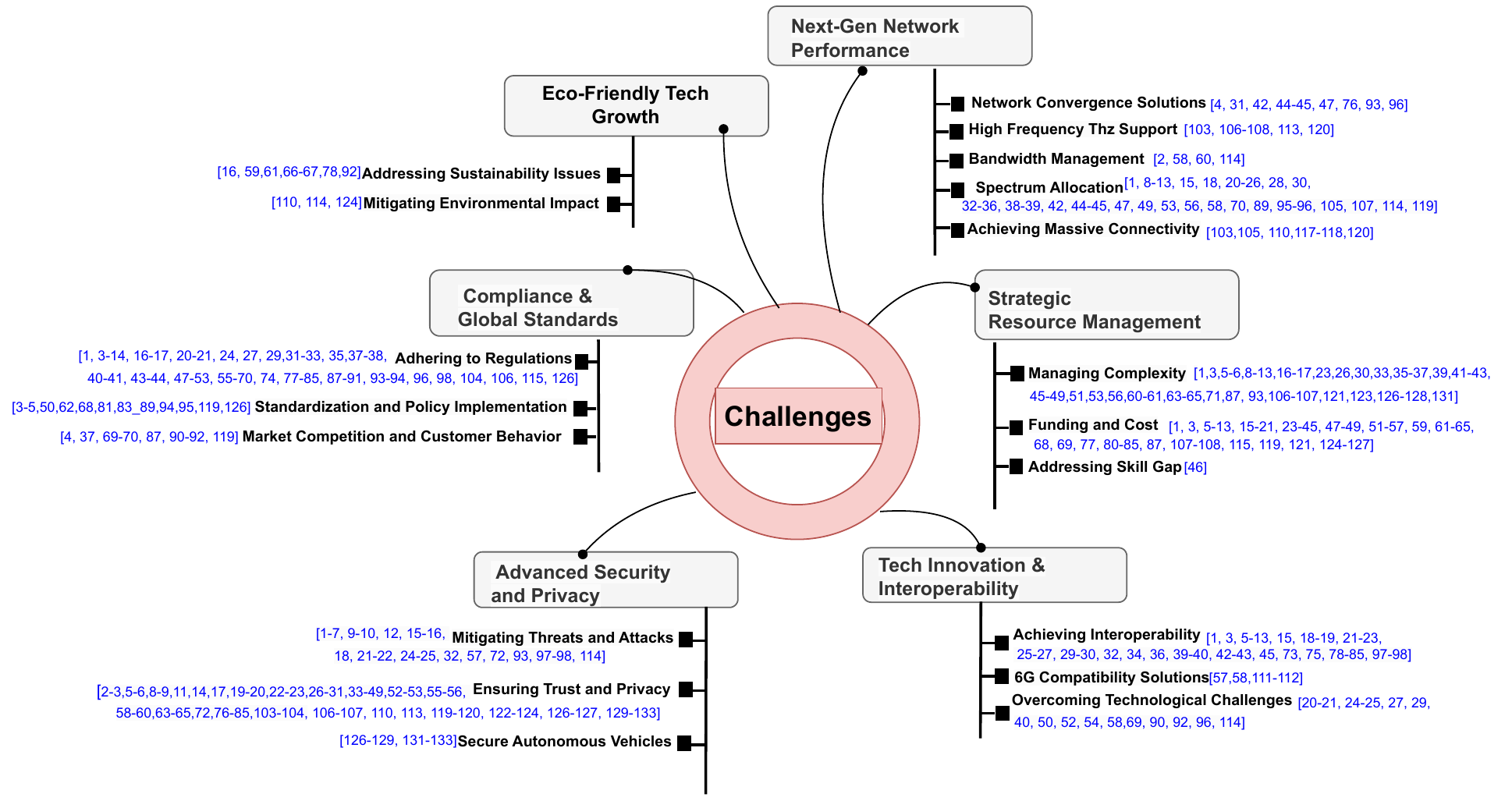}
  \caption{Core categories of 6G software challenges}
  \label{fig:chal}
\end{sidewaysfigure}

The 19 challenges identified for RQ3 were organized into six higher-order themes: Next-Generation Network Performance, Cybersecurity and Information Privacy, Compliance and Global Standards, Strategic Resource Management, Innovation and Integration, and Sustainability. Figure~\ref{fig:chal} summarizes these challenges across the six themes.

\subsubsection{Next Generation Network Performance: Scalability and Spectrum Limitations}

With 6G ecosystems encompassing a variety of different technologies and networks, (Chal1) Network Convergence, is a significant business opportunity and risk. Software companies need to integrate satellites and land/wireless networks in a single service offering, but poor inter-operability (gaps in inter-operability) and variable latency (variability in latency) impede seamless end-to-end performance (FL-04, FL-31, FL-42, FL-44--FL-45, FL-47, FL-76, FL-93, FL-96). Also, higher frequency bands have a signal (Chal2) high frequency THz support challenge, where there is high signal attenuation, short range (limited) transmission, and high deployment costs; these may limit scalable service deployment (FL-01, FL-10; GL-01, GL-04--GL-06, GL-11, GL-18). Concurrently, (Chal3) Explosive data growth is a significant issue, as it increases the need for efficient bandwidth management to avoid data congestion and service degradation; this is in addition to the need for sophisticated AI driven (adaptive) spectrum sharing and traffic management (FL-02, FL-58, FL-60; GL-12). Apart from the challenges of technical feasibility, (Chal4) spectrum allocation is a regulatory and business challenge due to the fragmented licensing of spectrum; this complicates cross industry and cross border service deployment (FL-01, FL-15, FL-18, FL-20--FL-26, FL-38--FL-39, FL-42, FL-47, FL-49; GL-03, GL-05, GL-12). Lastly, (Chal5) massive connectivity challenges are driven by the demand for support, the needs to connect billions of devices without negatively impacting service quality and cost; this places a continuous demand (active demand) for software-driven resource orchestration (FL-10, FL-24, FL-46; GL-01, GL-15--GL-16, GL-18).

\subsubsection{Heightened Threats to Safety and Confidentiality: Risks to Cybersecurity and Information Privacy}

The nature of 6G software ecosystems increases security-related business threats. (Chal6) Mitigating threats and attacks demand adaptable defenses to AI-based cyberattacks, quantum breaches, and adversarial manipulation, affecting trust, compliance, and service continuity (FL-09--FL-10, FL-12, FL-21--FL-22, FL-97--FL-98; GL-12). (Chal7) Trust and Privacy Solutions, as persistent challenges, are because of decentralized identity, AI-based authentication and data governance, and other frameworks must be integrated across multi-actor ecosystems. Without establishing trusted security frameworks, the adoption of platforms and the engagement in ecosystems are severely limited (FL-52--FL-53, FL-55--FL-56, FL-58--FL-60, FL-63--FL-65; GL-01--GL-02, GL-04--GL-05, GL-08, GL-17--GL-18, GL-20--GL-22). Sectoral specific threats are most apparent in (Chal8) protecting autonomous vehicles, where the shortcomings of vehicle-to-everything (V2X) communications and AI-based decision-making pose real safety, legal, and ethical risks to the reputations of software providers (FL-40, FL-42, FL-47; GL-24--GL-27, GL-29--GL-31).

\subsubsection{Compliance and Global Standards: Regulatory and Governance Challenges}

Apart from the technical uncertainties, (Chal9) regulatory compliance holds software for 6G businesses operating in multiple jurisdictions. The fragmented and conflicting international policies for spectrum allocation, governance of AI, and protection of personal data increase the cost of compliance and the possibilities of standard software solutions (FL-01, FL-16--FL-17, FL-20--FL-21, FL-40--FL-42, FL-74, FL-89; GL-04, GL-13). (Chal10) The gap between the speed of technological advances and the standardization and policy gap creates uncertainty regarding interoperability, certification, and investment in technologies for the long term (FL-50, FL-62, FL-68, FL-81, FL-95; GL-17, GL-24). From a business standpoint, uncertainty negatively impacts market timing, orchestration of ecosystems, and strategies around platform lock-in. Moreover, (Chal11) the market landscape and behavioral profile of customers engender adoption uncertainty wherein the demand, the elasticity of pricing, and the competitive landscape increase the complexity of revenue predictions and the stabilization of the business model (FL-04, FL-37, FL-69--FL-70, FL-87, FL-90--FL-92; GL-17).

\subsubsection{Strategic Resource Management: Economic and Operational Challenges}

6G infrastructure is highly capital intensive, resulting in (Chal12) cost and funding obstacles, especially for smaller companies and developing areas, creating unequal participation in the ecosystem and delaying market diffusion (FL-01, FL-47--FL-49, FL-51--FL-57, FL-59, FL-61--FL-65, FL-87; GL-05--GL-06, GL-19). From an operational perspective, (Chal13) complexity management is difficult as software companies must manage decentralized and highly autonomous and interdependent AI infrastructures (FL-13, FL-16--FL-17, FL-47, FL-60--FL-61, FL-63--FL-65, FL-87; GL-04--GL-05, GL-24--GL-26). This complexity increases operational risk and reliance on higher levels of automation. In addition, (Chal14) specialists in AI, quantum technology, and cybersecurity may be difficult to recruit, creating strategic challenges for the workforce planning and talent development of 6G software businesses (FL-01, FL-10, FL-46).

\subsubsection{Innovation and Issues of Integration}

In the realm of proven and emergent systems, (Chal15) interoperability has been a continuous issue, especially with regards to backward compatibility, service continuity, and integration among the ecosystem (FL-01, FL-15, FL-18, FL-21, FL-25--FL-27, FL-39--FL-40, FL-75). This means that (Chal16) solutions that support 6G compatibility requires a lot of testing, aligning of standards and certification thus prolonging the development cycles and increasing the operations cost of software enterprises (FL-52, FL-58; GL-09--GL-10). Besides, (Chal17) limitations, such as signal degradation, interference and inefficient energy consumption, still hamper scalable and economically viable deployment (FL-03, FL-24--FL-27, FL-29, FL-40, FL-52, FL-58, FL-69, FL-90, FL-92, FL-96; GL-12).

\subsubsection{The Growth of Eco-Friendly Technology: The Restrictions of Sustainability}

Sustainability has also become a cross-cutting constraint of 6G software ecosystems. To decrease the (Chal18) environmental impact, energy-efficient architecture, green networking strategies, and optimized AI models should be implemented to lower the carbon footprint of large-scale deployments (FL-16, FL-59, FL-61, FL-66--FL-67, FL-78, FL-92). In addition to the efficiency of operations, the (Chal19) sustainability concerns also require the practices of circular economies, responsible hardware life cycles, and long-term energy governance that, over time, are having an increasing impact on the regulatory compliance, the popular acceptance, and the strategic legitimacy of 6G software business (GL-08, GL-12, GL-22).

Overall, the identified challenges constrain the ability of 6G software businesses to scale services, coordinate ecosystems, and translate technological capabilities into sustainable business value.

\par\medskip
\noindent\rule{\columnwidth}{0.4pt}
\vspace{3pt}

\noindent\textbf{Key Takeaways (RQ3)}

\begin{enumerate}
    \item Network convergence, high-frequency operation, data growth, spectrum allocation, and massive connectivity create interrelated scalability and performance constraints for 6G software services.

    \item Cybersecurity threats, trust and privacy requirements, and safety risks in autonomous environments can affect service continuity, adoption, and confidence in software-enabled 6G services.

    \item Regulatory fragmentation, standardization gaps, and uncertain customer and market behavior create governance and commercialization challenges.

    \item High costs, operational complexity, and shortages of specialized skills constrain organizational readiness and participation in emerging 6G ecosystems.

    \item Interoperability, compatibility, technological limitations, and sustainability requirements indicate that scalable 6G software businesses depend on both technical integration and environmentally sustainable operation.
\end{enumerate}

\vspace{1pt}
\noindent\rule{\columnwidth}{0.4pt}
\par\medskip
\subsection{Framework Synthesis: The Integrative Perspective of 6G
Software
Business}\label{sec:cross-rq-synthesis}

A bottom-up synthesis approach, based on thematic analysis of the 133
primary sources, was used to derive the framework. In the framework, the
value that 6G software businesses create, constrain, and realize is
integrated. The framework integrates the characteristics (RQ1),
opportunities (RQ2), and challenges (RQ3) into a multi-factor framework,
which is a direct outcome from the analysis of themes from section
4.2-4.4. The themes were mapped to each layer based on these dependency
patterns. Each layer corresponds to a different theme. This separation
into layers represents not technical architecture, but a
business-relevant causal structure. This separation clarifies how value
is created in 6G software businesses through interactions rather than
through capabilities in isolation. The relationship between themes and
layers has been documented in the excel sheet in the supplementary
materials (Supplementary Material S1), which supports reproducibility
and auditability. It aims to facilitate structured reasoning regarding
6G software businesses in pre-standardization scenarios, where the
technological, business, and regulatory dimensions are still evolving in
conjunction.

\subsubsection{Layer 1: Business Context and Value
Drivers}\label{layer-1-business-context-and-value-drivers}

The first layer reflects the business context which comprises external
and strategic elements that shape the behavior of software businesses
within the 6G ecosystems. This layer comprises the themes of market
structure, regulation, sustainability, and the configuration of
ecosystems which have been described as boundary conditions throughout
literature.

These value drivers, on the other hand, do not create value on their
own. They outline the types of value that may be present and the types
that may be absent. For instance, in the case of sustainability,
regulation, and the energy efficiency gap, there are a few software
solutions that are socially and economically feasible; there are also
some solutions that are not economically and socially feasible, but they
do not conform to the sustainability regulation. On the other hand, the
tiered fragmentation of regulation determines the market access route.
This layer sets the context against which software business decisions
are made.

\subsubsection{Layer 2: Software
Characteristics}\label{layer-2-software-characteristics}

The second layer includes software characteristics from RQ1. The
characteristics reflect capability patterns tied to the 6G operational
landscape and software business. These include AI automation, real-time
analytics, agile resource allocation, manageability, interoperability,
and orchestrated intelligent networks.

Most importantly, these characteristics represent business capabilities,
and not technologies. For example, the business implications of
ultra-low latency are interpreted from the perspectives of service
responsiveness and competitive differentiation, and AI integration is
framed as a driver of operational scalability and cost efficiency. This
abstraction shifts the focus from technological novelty to
organizational and economic implications.

\subsubsection{Layer 3: Opportunity
Mechanisms}\label{layer-3-opportunity-mechanisms}

For the third layer, the focus is placed on the opportunity mechanisms,
as captured in RQ2. This layer focuses on how enabling characteristics
can be translated into value creation and capture. This includes
platform-based business models, AI-enabled services, vertical-specific
solutions, trust-based offerings, and digital services with immersive
experience. In this layer, specific opportunities correspond to specific
enablers. For instance, ecosystem orchestration opportunities require
interoperability and software architectures that can scale. In contrast,
services based on trust require automated security and decentralized
identity management. Such a dependency structure explains why
opportunities must be pursued in an interconnected manner, emphasizing
again the integrative characteristics of the framework.

\subsubsection{Layer 4: Cross-Cutting Challenges and
Constraints}\label{layer-4-cross-cutting-challenges-and-constraints}

Layer four captures the cross-cutting constraints identified in RQ3,
including security, regulation, cost, skills, sustainability, and
interoperability. These cross-cutting challenges limit the deployment of
capabilities and realization of opportunities and span across the other
layers, horizontally. Instead of viewing challenges in isolation, the
framework reframes each of the challenges as a structural friction
point. For example, the costs of infrastructure hold back scalability
even when the enablers are present, while the uncertainty of governance
can cause delays in deploying an ecosystem-level platform strategy. This
explains why the technological advancements have left an abundance of
unrealized 6G opportunities.

In summary, the integrative framework provides a more structured,
traceable and business-oriented consolidation of the disparate 6G
software ecosystems literature. It shows that for the 6G value to be
actualized, there must be a convergence of the context, software
capabilities, opportunity mechanisms and the management of constraints.
This response to the research objectives lays the groundwork for the
subsequent discussion and implications that will be elaborated in the
ensuing sections.

\section{Discussion}
\label{sec:discussion}

This section describes the synthesized distinct nature, challenges, and
prospects of the literature, using the integrated framework reviewed in
Section 4.5. While the results section organized these findings in the
form of a narrative and addressed research questions, this discussion
examines how these aspects collectively constitute 6G software
businesses as socio-technical and socio-economic systems. A primary
synthesis insight is that the value that is generated in 6G software
businesses is neither certain, nor unconditional. New business
activities can be generated due to the enhanced software functionalities
of intelligent automation, real-time processing, and scalable
architecture. However, this is only possible in the context of how
systemic constraints such as regulatory, security, cost, and
sustainability, challenges are faced. This clarifies the reason most
previous studies provide disconnected and inconsistent insights when
analyzing technological facilitators, enablers, and business models in
isolation.

\subsection{Interdependence of Capabilities, Constraints, and
Value}\label{interdependence-of-capabilities-constraints-and-value}

The framework shows that enabling features, challenges, and
opportunities are not disconnected elements. Rather, they form
interrelated capability-constraint-value relationships that collectively
shape the business's potential. These elements provide the necessary
technical and organizational capabilities. Challenges act as enduring
cross-cutting tensions, and opportunities are conditional pathways
instead of certain outcomes. This interdependence challenges the linear
narratives found in the 6G literature where business success is assumed
to follow from technical progress. The synthesis shows that
organizational maturity, ecosystem structure, and regulatory context may
vary even when capabilities are similar. By providing a more realistic
picture of pre-standardization environments, the framework shows how 6G
software businesses interact.

\subsection{Business Value Drivers as an Integrative
Lens}\label{business-value-drivers-as-an-integrative-lens}

The literature review identifies four business value drivers that are
recurrent across characteristics, opportunities, and challenges of the
reviewed literature. These value drivers, business goals, cost
efficiency, innovation and differentiation, market and ecosystem value,
and sustainability, are not introduced as goals to be accomplished, but
rather emerge as value based strategic orientation that influences the
decision-making process in 6G software enterprises.

The ability to control costs provides the foundation for scale and the
economic viability for sustainability in environments with high
complexity of infrastructure and operations. Intelligent and autonomous
capabilities create new avenues for innovation and differentiation but
also pose new governance and interoperability risks. The value of the
market and the ecosystem demonstrate the shift to platform-oriented and
multi-actor business models, which are dependent on trust and the
coordination of actors. The impact of sustainability on the design and
operation of systems increasingly constrains the long-term viability of
systems with respect to energy, regulatory acceptance, and
socio-political acceptance. These factors are interconnected in a
dynamic fashion rather than operating in isolation. For instance,
differentiation through innovation may lead to greater cost and
regulatory risks, while sustainability may diminish short-term
efficiency but increase long-term legitimacy of the ecosystem. The
framework represents the trade-offs, thereby enabling more complex
reasoning than unidimensional approaches.

\subsection{Conceptual Contribution Beyond Technical
Narratives}\label{conceptual-contribution-beyond-technical-narratives}

The chief conceptual contribution of this research is the repositioning
of 6G software businesses as actors that are embedded in ecosystems
rather than functioning as solely technical facilitators. While most of
the literature focuses on performance, latency, or novel architecture,
the framework shows that value creation is as much a function of
organizational integration and governance as it is of ecosystem
positioning.

The framework provides an integrated approach to description. It does
not specify maturity levels, readiness assessments, or ideal paths to
adoption. Rather, it assists in consolidating disparate pieces of
knowledge into an integrated framework that supports sensemaking across
the technical, business, and policy domains and supports the
construction of empirical synthesis.

\begin{figure}[p]
  \centering
  \includegraphics[width=\textwidth,height=0.88\textheight,keepaspectratio]{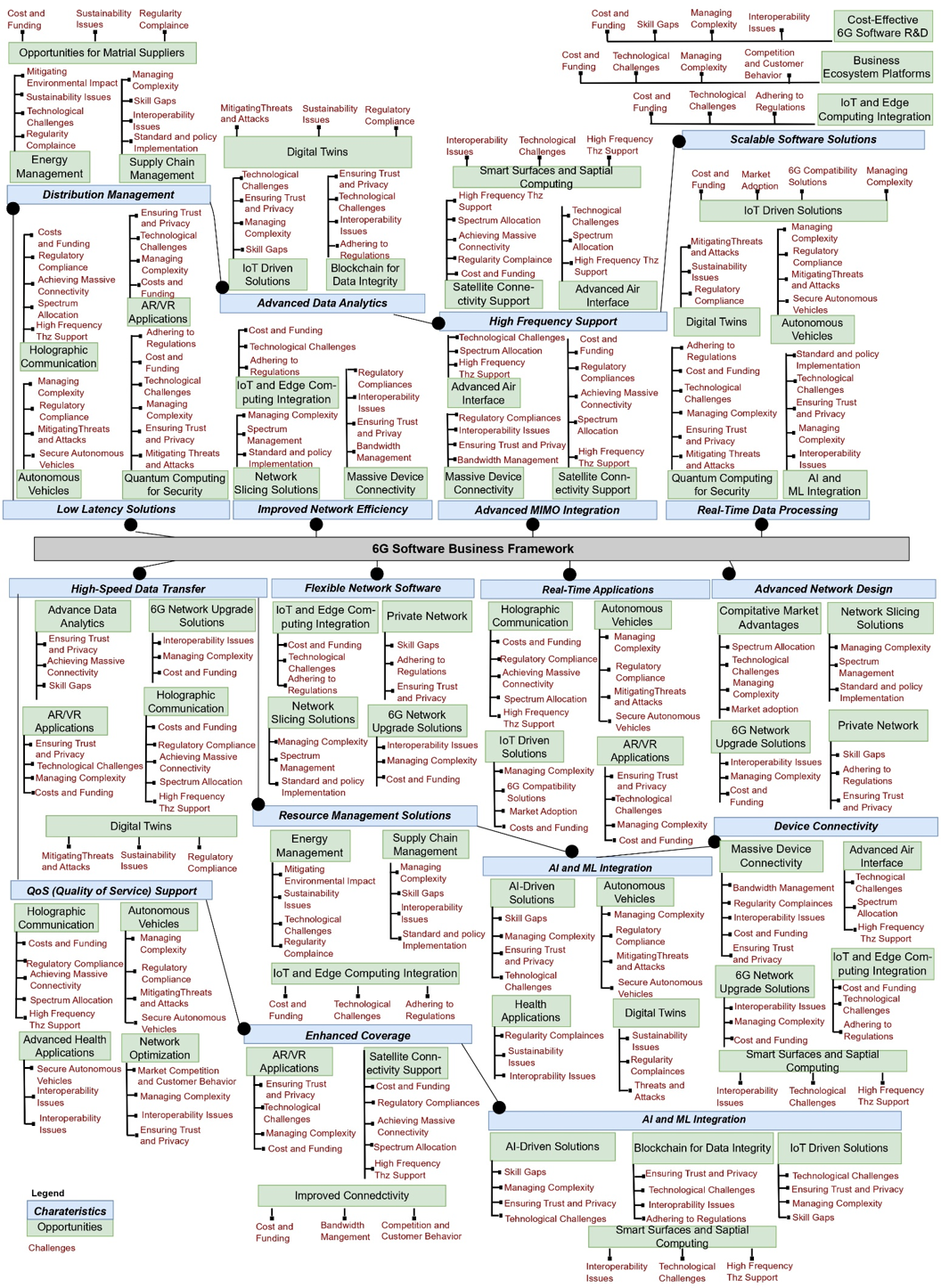}
  \caption{6G Software Business Framework. The figure synthesizes the characteristics, opportunities, and challenges identified in the review.}
  \label{fig:framework}
\end{figure}

\subsection{Business Ecosystem Model: From Isolated Technologies to
Integrated
Networks}\label{business-ecosystem-model-from-isolated-technologies-to-integrated-networks}

The 6G business landscape is moving away from a product-centric solution
towards ecosystem driven solutions. This framework reveals that 6G
success is dependent upon multi stakeholder collaboration, where
different industries will need to collaborate to achieve an effortless
adoption \textnormal{[83]}. It is no longer possible for companies to
build solutions alone; instead, they must work with cooperative networks
that leverage IoT, blockchain, AI, and edge computing-driven solutions
by different providers \textnormal{[71]}. For instance, 6G autonomous
vehicle development requires telecom providers, AI firms, cybersecurity
experts, and governmental regulators to collaborate on a standardized
connectivity and safety framework \textnormal{[71,87]}. Also, 6G
business models may shift towards decentralized governance rather than
traditional, centralized approaches. This decentralization raises new
challenges in interoperability, standardization, and trust management,
requiring a balance between decentralized security models and regulatory
supervision \textnormal{[82,95]}. This framework shows an
ecosystem-based approach, where different industries and technological
enablers need to be working together to form an interconnected entity
rather than as independent sectors.

\subsection{Emerging Trends: Strategic Shifts in 6G Software Business
Development}\label{emerging-trends-strategic-shifts-in-6g-software-business-development}

The framework reflects trends that will change how 6G software
businesses. Below are four key strategic shifts that redefine the
evolving landscape 6G technology:

\textbf{a.}~~~~~ \textbf{Integration of AI and edge computing: }The
integration of AI and edge computing enables real-time decision-making
at the device level by reducing the reliance on centralized cloud
models. This shift enhances data privacy and low-latency computing;
however, it also introduces security risks due to decentralized attack
surfaces.

\textbf{b.}~~~~ \textbf{Quantum Computing for Next-Generation Security:
}The structure framework shows that quantum computing will be necessary
for cryptographic security in 6G business models, as threats and attacks
grow. However, quantum security also results in energy consumption
issues, which need to be addressed by energy efficient quantum
processing units (QPUs).

\textbf{c.}~~~~~ \textbf{Outcome-Based Revenue Model:} Traditional
subscription-based telecom models are evolving into outcome-driven
monetization strategies. Instead of charging only for bandwidth usage,
6G businesses will generate revenue from data-driven intelligence,
network efficiency, and AI-enhanced insights. Example: Rather than
selling bandwidth as a service, 6G business models will leverage
AI-driven revenue streams, such as predictive analytics, automated
network management, and cybersecurity threat detection.

\textbf{d.}~~~~ \textbf{Sustainable energy management:} Energy
management is listed as a key challenge and opportunity in the taxonomy,
and 6G businesses must embed sustainability in their core strategy. In
6G enterprises, competitive advantage will be defined by AI optimized
energy consumption, green data centers and sustainable spectrum
management. The trends suggest that the 6G software businesses are
moving from traditional telecom services to AI driven, energy conscious,
decentralized and intelligence-based business models.

These emerging trends indicate that 6G software businesses are
transitioning from traditional telecom services to AI-driven,
energy-conscious, decentralized, and intelligence-based business models.

\section{Threats to Validity}
\label{sec:validity}

In this section, we provide an overview of the potential threats to
validity the study faces, and the strategies employed to minimize their
impacts. Following the study's MLR and evidence-based software
engineering principles, we assess threats within the framework of
construct validity, internal validity, external validity, and
reliability \cite{ref17,ref46,ref52}.

\subsection{Construct Validity}\label{construct-validity}

Construct validity assesses whether the study accurately describes the
characteristics, opportunities and challenges, and boundaries of the
software businesses of 6G.

The most significant threat is the absence of the standard definition of
the term ``6G software business,'' and, therefore, the lack of a
universally accepted definition of the term within academic and
practitioner literature creates a definitional inconsistency of the term
`6G software business.' Literature showing definitional inconsistency
and lack of a standard definition of a term is common within nascent and
pre-standardization fields and has the tendency to lead to conceptual
ambiguity \cite{ref17,ref46}. In this instance, I employed an
inductive synthesis process whereby the conceptual definition stems from
the observed recurring phenomena within the literature rather than a
priori definitional framework. Such an approach is consistent with the
recommendations provided for thematic synthesis within exploratory MLRs
\cite{ref17,ref61,ref62}.The use of grey literature presents
additional risks to the construct validity of the study due to the
variability of terms used, levels of analysis, and the depth and rigor
of analyses and reporting done in grey literature
\cite{ref17,ref58}. These risks were reduced by the use of explicit
inclusion criteria, the assessment of the inclusion and exclusion
criteria, the assessment of the thematic coding and the rules requiring
each theme to be substantiated by multiple sources. In accordance with
the established MLR guidance \cite{ref17,ref52}, grey literature
was used to complement, rather than substitute, the peer-reviewed
literature.

\subsection{Internal Validity}\label{internal-validity}

Internal validity is the extent to which the findings and the associated
framework logically follow from the data that were analyzed. Thematic
analysis is interpretive in nature, and therefore, the potential for
researcher bias, particularly in the coding and synthesis of the data,
poses a potential threat \cite{ref61,ref62}. This threat was
addressed by the implementation of collaborative coding among several
authors, an independent thematic review, and a consensus-building
activity to resolve any outstanding issues among authors, as prescribed
by the qualitative synthesis guidelines \cite{ref52,ref61}.The
derivation of the integrative framework is also a potential threat, as
some may argue that it is overly subjective. For this reason, the
various components of the framework were tied to the three research
questions, and throughout the analysis, efforts were made to maintain
the relationships among the raw data, themes, and the various
stratifications of the framework \cite{ref17,ref46}. This
supportive analysis was provided in the supplementary dataset to enhance
transparency and to demonstrate the auditability of the analysis.

\subsection{External Validity}\label{external-validity}

The external validity of a study is concerned with the extent to which
results can be generalized. Due to the fact that most of the literature
on 6G is conceptual, exploratory, or anticipatory because the technology
is still in the pre-standards and pre-deployment phase, statistical
generalization, is neither feasible nor applicable in this context
\cite{ref46}. This study focuses on analytical generalizations based
on the identification of patterns from differing sources, both academic
and practicing, that increase in diversity, or in this case, provide
enough diversity \cite{ref17,ref52}.~ Although grey literature
increases the practical relevance of the framework, it is mostly the
result of a specific region, organization, or industry, and therefore,
it is appropriate to view the framework, in this instance, as a
descriptive and integrative sensemaking framework, rather than as a
prescriptive or predictive framework. The legacy of establishing
empirical evidence -- through case studies, surveys, or other means --
is a fundamental legacy that should be preserved for future research.

\subsection{Reliability}\label{reliability}

How the research is conducted and the processes that are followed are
some of the things that are considered when attempting to address the
reliability of the study.

A possible threat in this area is the construction of a search string
that is specific to the database and that the construction of the search
string has been modified in order to account for the differential
indexing, query construction, and retrieval mechanisms of the various
digital libraries. This sort of thing usually happens when conducting
systematic and multivocal reviews, and the lack of reliability
associated with this sort of thing is not the result of lack of
reliability, but rather a case in which reporting is not systematic
\cite{ref17,ref46,ref52}. The threat was lessened by the fact
that all search strategies were documented, the reasons for going with a
specific database were articulated, and a uniform inclusion and
exclusion criterion was implemented across the various sources. The
analytical reliability has been enhanced by evaluating inter-rater
agreement using Kendall's coefficient of concordance, demonstrating
consistency across evaluators \cite{ref18}. Moreover, as a matter of
transparency, replicability, and the possibility of conducting secondary
analyses, evaluators maintained full access to the entire dataset, the
quality assessment scores, and the coding schema as supplementary
materials.

The study exhibited and utilized procedural safeguards to reduce the
impacts of the limitations tied to the conceptual maturity, data
heterogeneity, and/or the interpretive synthesis of the final product of
an MLR in an emergent area of study \cite{ref17,ref46}. The
credibility and rigor of the synthesis are fostered by the combination
of procedural transparency, collaborative peer review, analytical
reliability assessments, and traceability.

\section{Study Implications and Future
Work}
\label{sec:implications}

The integrative framework constructed in this study has both practical
and research implications by offering a framework for reasoning
regarding 6G software businesses from a practice-oriented technological,
organizational, and regulatory perspective. The framework does not
provide a prescriptive model or a means for conducting a maturity
assessment. Rather, it brings together disparate pieces of information
into a single framework for guiding decision-making and facilitating
systematic research.

From a practical standpoint, the framework contributes to
interpretations of 6G software initiatives that go beyond individual
technological components or use cases. The framework allows for the
identification of enabling components, systemic barriers, and mechanisms
to capture value, and this contributes to the development of real-world
software-centric capabilities that can be transformed into business
value. Based on the framework, \textbf{practitioners} can determine
whether the 6G initiatives proposed are aligned with the primary
business value objectives (cost savings, innovation, participation in
ecosystems, and sustainability) identified in the research. This
prevents organizationally unprepared, governance absent, or
sustainability overlooked technological decision-making. The framework
emphasizes that value realization for 6G software companies is
inherently conditional. Intelligent automation, immersive apps, or
ecosystem-based platforms have value, but due to regulatory uncertainty,
security risks, interoperability issues, and sustainability challenges
the realization of value is conditional. \textbf{For practitioners,}
this entails adopting strategies that are incremental and contextual,
rather than relying on predictions of linear technological development.
With the advent of agentic AI, autonomously governed 6G systems will
result in an increased complexity of governance, and the framework
encourages the proactive management of cross-cutting risks and the early
trade-offs that impact design and organizational architecture.

The framework illustrates an empirically grounded artifact that
\textbf{researchers} can use in their operationalization of the
framework in their research. Each component of the framework is the
result of the multivocal synthesis, documented in the supplementary
dataset and provides a clear basis for designing interview protocols,
surveys, and case study frameworks. The framework \textbf{helps
researchers} go beyond the mere description of the business value
drivers and operational constraints in the case of different
organizations, or the abstraction and recombination of capabilities, to
evaluative and explanatory research.

The implication is based on research around agentic AI and
software-intensive ecosystems. The other highlighted features of
real-time intelligence, autonomous decision-making, and ecosystem
coordination are also seen in the evolving research on agentic AI. This
framework places these features in the context of \textbf{business and
governance} which allows for the study of agentic AI as not just a
technical feature, but also aids in locating where governance tools,
like the enforcement of standards, regulatory alignment, or
subsidization for sustainable practices, can mitigate obstacles to the
realization of value \textbf{for policymakers}.

The study contributes a structure for understanding the field rather
than proposing specific solutions. The framework, by being explicit on
the interplay between the three dimensions of capability, constraint,
and value, offers a point of reference for systematic reasoning to the
\textbf{researchers, practitioners, and policymakers} on the
software-centric business model evolution in the anticipated 6G
ecosystems.

\section{Conclusion}
\label{sec:conclusion}

In this study, the authors provide a systematic assessment of the
nascent 6G software industry, using a multivocal approach literature
review that integrates literature from both academia and industry
professionals. By compiling 133 sources of literature, this study was
able to answer 3 questions describing the characteristics, challenges,
and opportunities, that would shape software-centric value systems in
the pre-standardization 6G ecosystems. The findings suggest that the
defining factors of 6G software businesses are not singular, but rather
in intersecting patterns of capabilities such as intelligent automation,
real-time analytics, scalable software architecture, and the ability to
be interoperable to a wide variety of systems. Additionally, the
findings provide evidence of pervasive security issues, fragmented
regulations, operational complexity, and sustainability that would limit
the ability for these capabilities to be transposed to a viable source
of business value. The identified challenges provide evidence supporting
the existence of business value, created and captured in a variety of
ways, through trust-enabled platforms and immersive applications,
industry-specific digital systems, and business ecosystems. Insights
have been synthesized into an integrative framework describing business
value drivers, enabling software features, opportunity pathways, and
challenges that constrain opportunities. The framework is descriptive in
nature and provides a structured view to help understand the emergence
of software-centric value in 6G, how it might be constrained and is
based on empirical thematic synthesis.

From a practical standpoint, the framework helps software producers,
network operators, and ecosystem players to pinpoint where they should
prioritize strategic investments, and where coordinated efforts are most
needed. For policymakers, the findings underscore the need to steer
regulatory and governance frameworks to align with the emerging
intelligence of networks that depend on software rather than traditional
hardware-centric thinking. For researchers, this study brings together
disparate segments of knowledge and provides a basis for constructing
theory and pursuing empirical validation.

This study is not free from the limitations of multivocal reviews,
including the reliance on existing literature and the pre-standards
interpretations of 6G. Thematic transparency, strict selection criteria,
and inter-rater reliability assessments strengthen the synthesis. Future
studies should focus on the empirical validation of the proposed
framework through expert assessments and practical case studies and
examine how business software configurations evolve with the 6G
standards. The research contributes to the comprehension of 6G software
enterprises by combining technological potential, entrepreneurial
prospects, and structural limitations into a singular conceptual
framework, providing a solid basis for future research and practical
efforts for next-generation software-centric network ecosystems.

\backmatter

\section*{Declarations}

\subsection*{CRediT Authorship Contribution Statement}

\textit{Muhammad Zohaib}: Writing--original draft, Methodology, Formal analysis. \textit{Muhammad Azeem Akbar}: Writing--review \& editing, Validation, Supervision. \textit{Sami Hyrynsalmi :} Writing--review \& editing, Supervision. \textit{Muhammad Hamza:} Writing--review \& editing.

\subsection*{Funding}
This research was funded by the Business Finland project 6G Bridge--6G software for extremely distributed and heterogeneous massive networks of connected devices (8516/31/2022).

\subsection*{Competing interests}
The authors declare no conflict of interest.

\subsection*{Data availability}
The data supporting this study are publicly available in the accompanying replication package on Zenodo at \url{https://doi.org/10.5281/zenodo.22132234}. The package includes the selected-study dataset, quality-assessment results, extracted and coded data, study-to-factor and factor-to-theme mappings, independent rater classifications, and inter-rater reliability calculations.

\subsection*{Declaration of generative AI and AI-assisted technologies}
During manuscript preparation and revision, the authors used OpenAI ChatGPT and Grammarly to assist with language editing, readability, reduction of repetition, and editorial refinement. ChatGPT-assisted editorial revisions concerned the Introduction, Background, Methodology, Results and Findings, Discussion, Related Work and Comparative Analysis, Threats to Validity, and Conclusion. The authors reviewed and revised the assisted material and take full responsibility for the research methods, evidence, citations, interpretations, and final manuscript. This statement concerns manuscript preparation and does not assert AI use in study screening, data extraction, or thematic coding.

\bibliography{references}

\end{document}